\documentclass[aps,prl,twocolumn,superscriptaddress,longbibliography,nobalancelastpage]{revtex4-1}

\usepackage{graphicx}
\usepackage{amsmath}
\usepackage{color}
\usepackage[normalem]{ulem}
\usepackage[utf8]{inputenc}
\usepackage{hyperref}
\usepackage{float}
\usepackage{siunitx}

\begin{document}
	\title{Spatial and {T}emporal {C}oherence in {S}trongly {C}oupled {P}lasmonic Bose--Einstein {C}ondensates}
	
	\author{Antti J. Moilanen}
	\affiliation{Department of Applied Physics, Aalto University School of Science,
		P.O. Box 15100, Aalto, FI-00076, Finland}
	
	\author{Konstantinos S. Daskalakis}
	\affiliation{Department of Applied Physics, Aalto University School of Science,
		P.O. Box 15100, Aalto, FI-00076, Finland}
	\affiliation{Department of Mechanical and Materials Engineering, Turku University Faculty of Technology, Turku, FI-20014, Finland}

	\author{Jani M. Taskinen}
	\affiliation{Department of Applied Physics, Aalto University School of Science,
		P.O. Box 15100, Aalto, FI-00076, Finland}
	
	\author{Päivi Törmä}
	\email{paivi.torma@aalto.fi}
	\affiliation{Department of Applied Physics, Aalto University School of Science,
		P.O. Box 15100, Aalto, FI-00076, Finland}
	
	\date{\today}
	
	\begin{abstract}
		
		We report first-order spatial and temporal correlations in strongly coupled plasmonic Bose--Einstein condensates. The condensate is large, more than twenty times the spatial coherence length of the polaritons in the uncondensed system and hundred times the healing length, making plasmonic lattices an attractive platform for studying long-range spatial correlations in two dimensions (2D). We find that both spatial and temporal coherence display non-exponential decay; the results suggest power-law or stretched exponential behaviour with different exponents for spatial and temporal correlation decays.  
		
	\end{abstract}
	
	\maketitle
	
	Three-dimensional Bose-Einstein condensates (BECs) in thermal equilibrium exhibit long-range order of spatial correlations which, in principle, extend to infinity. In 2D, true long-range order is prohibited by thermal fluctuations~\cite{mermin_absence_1966,hohenberg_existence_1967,bloch_many-body_2008}. Nevertheless, it has been shown that quasi-long-range order may persist in equilibrium systems through the Berezinskii–Kosterlitz–Thouless (BKT) transition~\cite{kosterlitz_ordering_1973,berezinskii_1971}, and in non-equilibrium via the dynamical phase ordering of Kardar–Parisi–Zhang (KPZ)~\cite{kardar_dynamic_1986}. 
	Exciton-polaritons (photon--exciton quasi-particles) offer a platform for studying correlations of driven-dissipative condensates in 2D, yet the occurrence of long-range order in these systems has remained elusive. Typically the systems have been too small to give definitive answers about decay at long distances. We introduce strongly coupled plasmonic BECs as a system for studying correlations: we demonstrate a condensate of size more than twenty times the spatial coherence length of the polaritons and two orders of magnitude larger than the healing length. In previous reports of exciton-polariton condensates and photon BECs the ratio of condensate size to length scales such as de Broglie wavelength or healing length has been approximately a factor of ten at best~\cite{caputo_topological_2018,daskalakis_spatial_2015,Damm2017NatComm,marelic_spatiotemporal_2016}.
	We find decay of spatial and temporal coherence that is clearly non-exponential; the results are best described by power-law or stretched exponential decay with different exponents for spatial and temporal correlations.
	
	Various scenarios of long-range correlation decay have been predicted for 2D systems. In equilibrium BECs, thermal fluctuations give rise to vortices and anti-vortices which disrupt the long-range order. In the BKT transition, below a critical temperature, vortices and anti-vortices are paired such that their phases cancel out allowing algebraic decay of correlations, $g^{(1)}(x)\propto x^{-b}$. The BKT transition entails power-law decays of both spatial and temporal correlations with equal exponents $b_\mathrm{s}=b_\mathrm{t}\leq 0.25$~\cite{nelson_universal_1977}, although in the presence of drive and dissipation the exponents may differ as $b_\mathrm{s}=2b_\mathrm{t}$~\cite{comaron_non-equilibrium_2021,szymanska_mean-field_2007}. 
	In non-equilibrium condensates, the occurrence of the BKT transition has been theoretically both supported~\cite{dagvadorj_nonequilibrium_2015,comaron_non-equilibrium_2021} and refuted~\cite{altman_two-dimensional_2015}. Long-range phase ordering could be restored by the KPZ mechanism with correlations decaying as a stretched exponential~\cite{altman_two-dimensional_2015,ferrier_searching_2020,comaron_dynamical_2018}. A crossover between the KPZ dynamics and equilibrium-like BKT has been proposed, determined by the degree of anisotropy in the system~\cite{altman_two-dimensional_2015,zamora_tuning_2017}. Contrary to BEC, for a usual laser an exponential decay of temporal correlations is expected~\cite{schawlow_infrared_1958}.
	
	Semiconductor polariton condensates~\cite{kasprzak_bose-einstein_2006,daskalakis_nonlinear_2014,plumhof_room-temperature_2014} are typically far from equilibrium conditions due to the short lifetime of polaritons with respect to their thermalization time, however, a quasi-equilibrium 
	state can be achieved by a dynamical balance of pump and dissipation~\cite{carusotto_quantum_2013,keeling_boseeinstein_2020}; near-equilibrium conditions are also possible~\cite{caputo_topological_2018,sun_bose-einstein_2017}. Polaritons decay via emission of photons, providing optical access to the properties of the condensate. This makes polariton condensates an attractive platform for studying spatial and temporal correlations. Early reports on spatial correlations in microcavity polariton condensates indicated algebraic decay~\cite{roumpos_power-law_2012, nitsche_algebraic_2014}, whereas temporal correlations have been shown to decay exponentially or as a Gaussian~\cite{baboux_unstable_2018,love_intrinsic_2008,haug_temporal_2012,whittaker_coherence_2009,spano_build_2013,daskalakis_spatial_2015}. There have been few studies governing both spatial and temporal correlations, thus the question of whether true long-range order exists in non-equilibrium polariton condensates remains open. At equilibrium-like conditions, with polariton lifetime exceeding the other timescales of the system, a BKT transition has been suggested with both spatial and temporal correlations showing a power-law decay with $b<0.25$~\cite{caputo_topological_2018}. In general, establishing power-law behaviour quantitatively is challenging~\cite{clauset_power-law_2009}. 
	
	Spatial and temporal correlations have been studied also in other luminous condensates. In photon BECs~\cite{Damm2017NatComm,marelic_spatiotemporal_2016} and plasmonic polariton condensates (polariton lasers)~\cite{de_giorgi_interaction_2018}, correlations have shown exponential and Gaussian decays; the absence of long-range order has been attributed to small condensate size and finite-size effects~\cite{Damm2017NatComm} and to the presence of drive and dissipation~\cite{de_giorgi_interaction_2018}.
	Indeed, one of the central factors hindering the studies of long-range order in polariton and photon condensates has been the small system size~\cite{comaron_dynamical_2018,altman_two-dimensional_2015}. 
	
	Here, we introduce a plasmonic polariton BEC with long-range spatial correlations that extend to remarkably long distances~\footnote{The condensate size \SI{500}{\micro m} is 20 times the spatial coherence length of the polaritons in the uncondensed system (below the first threshold, \SI{22}{\micro m}~\cite{SM}) and 100 times the healing length $\xi_0 =\hbar/\sqrt{{2m_\mathrm{eff}gn}}=0.6$...\SI{6}{\micro m}, where $m_\mathrm{eff}=1e^{-7}-1e^{-5} m_\mathrm{e}$ and $gn=0.02$~eV~\cite{Vakevainen2020}.}. We present the first measurement and in-depth analysis of both spatial and temporal coherence in plasmonic BECs. Our results clearly show that the correlations differ both from a non-ordered phase and from a laser. The exponents extracted by power-law and stretched exponential fits hint to a scenario between quasi-equilibrium and non-equilibrium.

	\begin{figure}
		\vspace{-8pt}
		\centering
		\includegraphics[width=0.73\columnwidth]{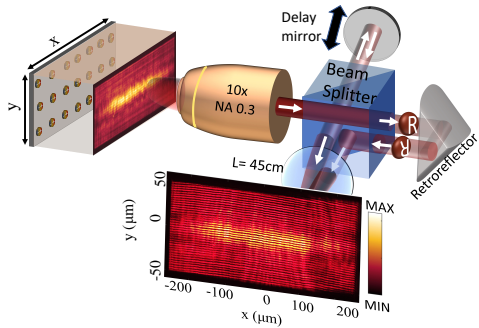}
		\vspace{-5pt}
		\caption{Schematic of the experiment. The sample luminescence is imaged through a microscope to a Michelson interferometer, where the real-space image is inverted over both $x$ and $y$ and overlapped with the original image.
		}
		\label{figure1}
		\vspace{-10pt}
	\end{figure}

	\textit{System and experiment.}---Plasmonic lattices, comprising nanoparticle arrays covered with fluorescent molecules, have been used to create BECs in the weak~\cite{hakala_bose-einstein_2018} and strong~\cite{Vakevainen2020} coupling regimes. Nanoparticle arrays give rise to surface lattice resonances (SLRs) which are hybrid modes of localized surface plasmon resonances in the individual nanoparticles and light diffracted to the periodic array~\cite{wang_rich_2018,kravets_plasmonic_2018,wang_structural_2018}. The $\Gamma$-point of the SLR dispersion provides a band edge for lasing~\cite{hakala_lasing_2017,daskalakis_ultrafast_2018} and condensation~\cite{Vakevainen2020,hakala_bose-einstein_2018}. To study spatial correlation decay at long distances, we use structures similar to those in our previous work~\cite{Vakevainen2020}, but extend the system in $x$ to \SI{500}{\micro m} while keeping the $y$ dimension at \SI{100}{\micro m}. The arrays are covered with an 80~mM solution of IR-792 dye leading to strong coupling between the molecules and the SLR modes, which persists also at high pump fluences~\cite{SM}. The plasmon-exciton-polaritons, called polaritons hereafter, in the system are bosonic quasi-particles consisting of light diffracted to the array, electron plasma oscillation, and the dye excitons. The molecules are excited by a pulsed laser (50~fs, 1~kHz, 800~nm/1.55~eV, 0.04eV FWHM) with a spot larger than the arrays. See~\cite{SM} for descriptions of the samples and the experimental setup.
	
\begin{figure*}
		\centering
		\includegraphics[width=0.82\textwidth]{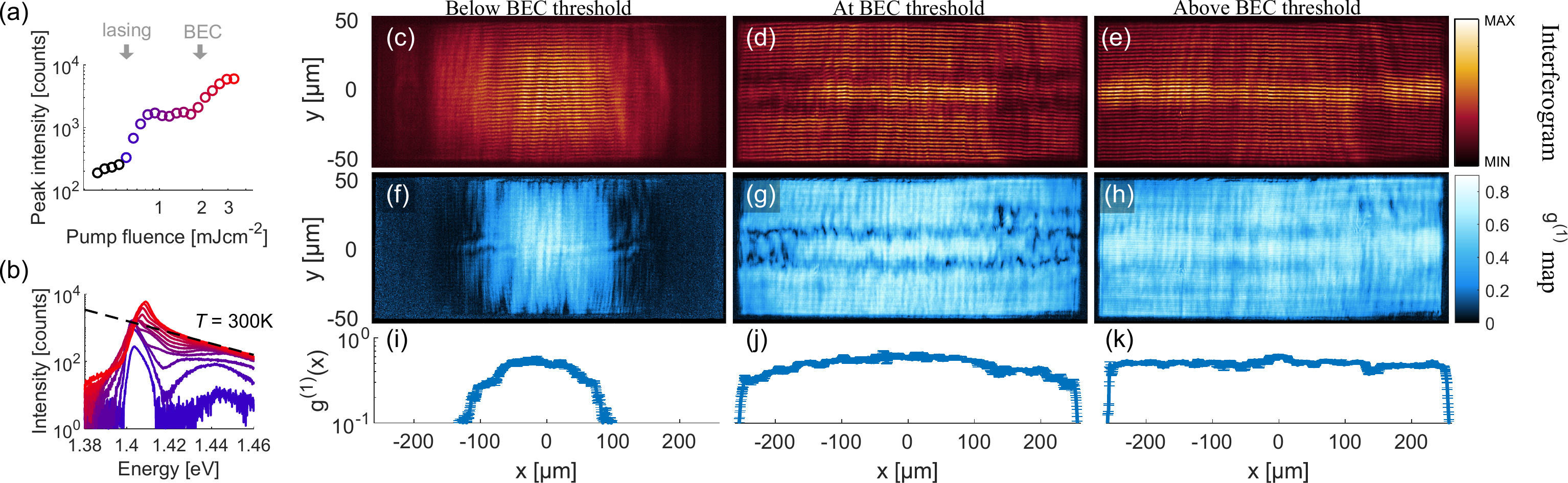}
		\vspace{-5pt}
		\caption{Interferograms and spatial correlation maps. (a) Threshold curve and (b) photoluminescence spectra. The pump fluences are color-coded from low (blue) to high (red). Interferograms at three pump fluences (c) below (0.83~mJcm$^{-2}$), (d) at (1.66~mJcm$^{-2}$), and (e) above (3.31~mJcm$^{-2}$) the BEC threshold. (f-h) Maps of $g^{(1)}(\textbf{r},-\textbf{r})$ for the corresponding pump fluences. (i-k) Average $g^{(1)}(x)$ over the $y$-axis of the array. The error bars show the standard deviation of three measurements. A sample with random distribution of nanoparticles did not show coherence (Fig.~S10~\cite{SM}). }
		\label{figure2}
		\vspace{-10pt}
	\end{figure*}

	As in our previous work~\cite{Vakevainen2020}, upon increasing pump fluence, the sample luminescence exhibits a double threshold, where the first threshold corresponds to polariton lasing and the second to polariton BEC (Fig.~\ref{figure2}(a)). Distinct from most polariton condensates, we observe a thermal distribution that extends over a range of 2$k_\mathrm{B}T$ at room temperature, see Fig.~\ref{figure2}(b). Since the polariton lifetime without gain $\sim$~100~fs is of the same order as the other timescales of the system, it is reasonable to assume the condensate to be in quasi- or non-equilibrium regime~\footnote{In this case, coherences may be present in the system density matrix even if the photon populations followed a thermal distribution.}. We observe a prominent change in the coherence properties between the lasing and BEC regimes.
	
	We measure spatial and temporal coherence by a Michelson interferometer in a mirror-retroreflector configuration, a schematic of the experiment is presented in Fig.~\ref{figure1}. The interference fringe contrast is directly proportional to the first-order correlation function between two points separated by $|\textbf{r}-\textbf{r'}|$, $g^{(1)}(\textbf{r}, \textbf{r}'; \tau)$, where $\tau$ is the time delay between the interferometer arms. Thus, changing the longitudinal position of the mirror and recording interferograms over a range of delays allows for directly probing the temporal coherence. For spatial coherence, we measure a series of interferograms at fixed intervals around $\tau=0$ over three periods of light frequency oscillation. The amplitude and phase of the interference fringes are obtained by fitting a sinusoidal function through each pixel of the stack of normalized interferograms; see~\cite{SM} for details.

	\textit{Spatial coherence.}---Figure~\ref{figure2} shows interferograms, the corresponding $g^{(1)}\left(\textbf{r}, -\textbf{r}; \tau=0 \right)$ maps, and average $g^{(1)}(x)$ along the long axis of the lattice for three excitation regimes: below, at, and above the BEC threshold. The phase maps of the fringes are presented in Fig.~S4~\cite{SM}. Below the threshold, spatial coherence emerges around the center of the array and decays towards the edges. At the threshold, coherence extends over a longer distance but gradual decay is visible. Remarkably, above the threshold, coherence is nearly constant throughout the array.
	
\begin{figure}[b]
		\vspace{-15pt}
		\centering
		\includegraphics[width=0.97\columnwidth]{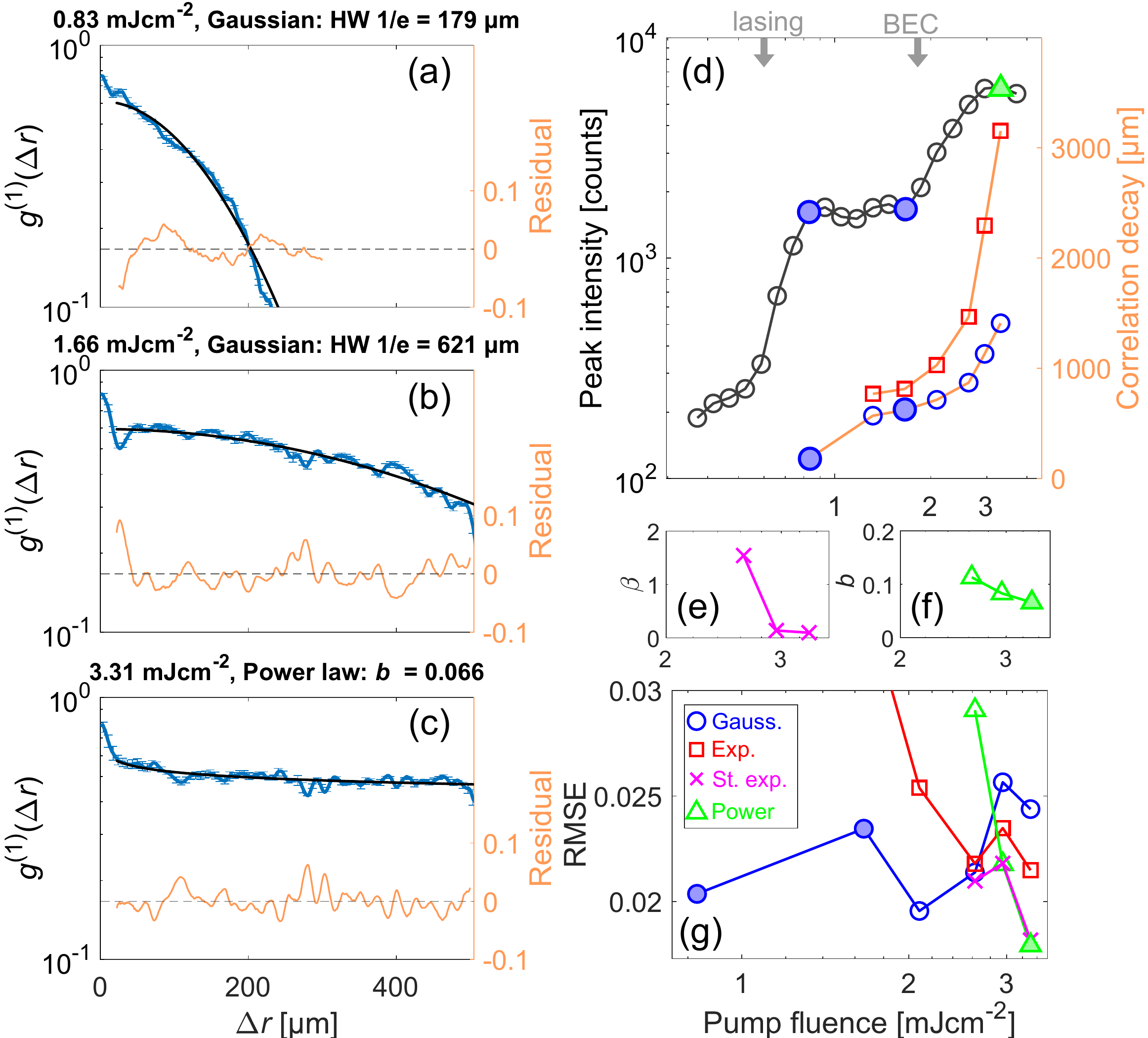}
		\caption{Quantitative analysis of spatial correlation decay. Measured $g^{(1)}(\Delta r)$ at three pump fluences (a) below, (b) at, and (c) above the BEC threshold. The error bars represent the standard deviation of three measurements. The best-fitting functions are shown by black solid lines. (d) Threshold curve (black circles) and spatial correlation decay extracted from Gaussian (blue circles) and exponential fits (red squares). The fluences corresponding to (a-c) are highlighted with filled symbols. For the exponential function, the reported length is the decay constant and for the Gaussian the half-width (HW) at 1/e decay. Exponents extracted from (e) stretched exponential and (f) power-law fits. (g)
			RMSE of the fits. In (d-f) the horizontal axis values are in the same units as in (g).}
		\label{figure3}
		\vspace{-10pt}
	\end{figure}

	Let us have a closer look at the decay of correlations. Fig.~\ref{figure3}(a-c) show $g^{(1)}$ as a function of radial separation of centrosymmetric points, $\Delta r=|\textbf{r}-\textbf{r'}|$, for the pump fluences highlighted with filled symbols in Fig.~\ref{figure3}(d). The measured $g^{(1)}(\Delta r)$ are fit to Gaussian, exponential, stretched exponential, and power-law functions: $g^{(1)}(\Delta r) = a ~\mathrm{exp}(-(\Delta r/d)^\beta)$ and $g^{(1)}(\Delta r) = a {(\Delta r)}^{-b}$,
	where $\beta=2$ gives a Gaussian and $\beta=1$ an exponential function, and $0<a\leq1$
	is a scaling parameter. Power-law or stretched exponential behaviour are expected to occur only above BEC threshold. We exclude the short-range regime around the autocorrelation point ($\Delta r=0$) from the fits, determined by the spatial coherence length of the polaritons in the uncondensed system (below the first threshold, \SI{22}{\micro m}~\cite{SM}). All the fits are performed using the same fit range, and the best-fitting models are shown in Fig.~\ref{figure3}(a-c). At pump fluences below the BEC threshold, as shown in Fig.~\ref{figure3}(a-b), spatial correlations decay as a Gaussian. Above the threshold, Fig.~\ref{figure3}(c), long-range spatial coherence covering the entire array emerges and the correlation function is nearly constant over a remarkably long distance: a fit to an exponential yields a decay length of more than \SI{3000}{\micro m}, which is six times the long axis of the system and \textit{two orders of magnitude} larger than the coherence length of the polaritons. The decay lengths given by Gaussian and exponential fits are presented in Fig.~\ref{figure3}(d) alongside the threshold curve. However, the data is best fit to a power-law function with a small exponent of 0.07, and almost equally to a stretched exponential with an exponent of around 0.1. The fits to all four functions are presented in Fig.~S8~\cite{SM}. Fig.~\ref{figure3}(e-f) show the exponents obtained from the stretched exponential and power-law fits. The root-mean-square error (RMSE) of the fits are compared in Fig.~\ref{figure3}(g).

	The thermalization in the plasmonic polariton BEC involves stimulated processes~\cite{Vakevainen2020}. Here, we separated the thermal tail (the part of the BEC spectrum which displays Maxwell-Boltzmann distribution) of the condensate emission and found that there seems to be some coherence present in the emission (Fig.~S7~\cite{SM}), suggesting such experiments can be used for future studies of the thermalization process. 
	
\begin{figure}[t]
		\centering
		\includegraphics[width=0.97\columnwidth]{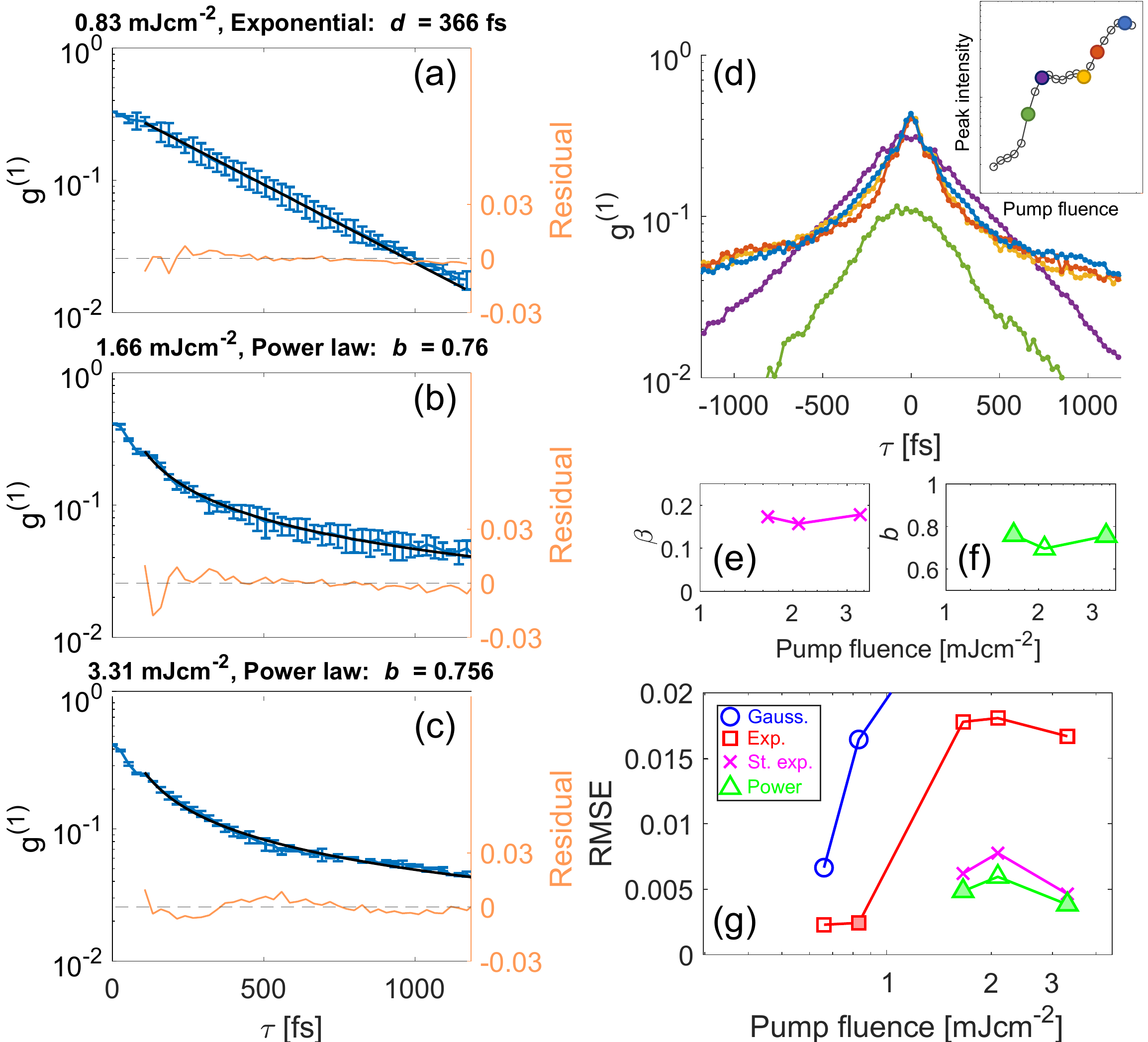}
		\caption{Quantitative analysis of temporal correlation decay. (a-c) Measured $g^{(1)}(|\tau|)$ at three pump fluences (a) below, (b) at, and (c) above the BEC threshold. The error bars represent the standard deviation of the averaged $\tau<0$ and $\tau>0$ values from (d). Fits to the best-fitting function are shown with black solid lines. (d) Measured temporal coherence for different pump fluences, color coded according to the filled circles in the threshold curve (inset). Exponents extracted from (e) stretched exponential and (f) power-law fits. (g) RMSE of the fits.}
		\label{figure4}
		\vspace{-10pt}
	\end{figure}

	\textit{Temporal coherence.}---Temporal coherence for increasing pump fluence is displayed in Fig.~\ref{figure4}(d). The measured $g^{(1)}(|\tau|)$ are again fit to Gaussian, exponential, stretched exponential, and power-law functions, and the best-fitting models are shown in Fig.~\ref{figure4}(a-c). The short-range correlations, between $\tau=0$ and the temporal coherence of the polaritons in the uncondensed system (below the first threshold, $\tau=104$~fs), are excluded from the fits~\cite{SM}. Below the BEC threshold, as shown in Fig.~\ref{figure4}(a), correlations decay exponentially with a decay constant of 366~fs. This already exceeds the temporal coherence of the polaritons due to the lasing triggered at the first threshold. Above the BEC threshold, Fig.~\ref{figure4}(b-c), the data is best fit to a power-law function with an exponent of around 0.7--0.8; the stretched exponential fits almost equally with an exponent of $\beta\sim0.2$, given by Fig.~\ref{figure4}(e-f). Fig.~\ref{figure4}(g) presents the RMSE of the fits. The decay of temporal coherence in the BEC regime is significantly slower than in the lasing regime (see also Fig.~S11~\cite{SM}). 
	
	The measured spatial coherence at $\Delta r$ is in general higher than the temporal one at $|\tau|=\Delta r/c$ ($c$ is the speed of light), $g^{(1)}(\Delta r) >> g^{(1)}(\Delta r/c)$, for instance for $\Delta r =$ \SI{300}{\micro m} ten times higher. This raises a concern about causality, needed for spontaneous formation of spatial coherence. However, this difference might be reduced by taking into account the different value of the two coherences at zero distance and time, and the finite photoluminescence duration, Fig.~S12~\cite{SM}. Based on Fig.~S12, the duration of the output pulse is likely to be about 2~ps or longer.
	
	\textit{Discussion.}---At pump fluences far below the BEC threshold but above the lasing threshold, the system behaves much like a polariton laser: spatial correlations decay as a Gaussian and temporal correlations as an exponential function. Nanoparticle array lasers at weak coupling regime have previously shown Gaussian decay of both spatial and temporal correlations~\cite{hoang_millimeter-scale_2017}, and at strong coupling an exponential decay of spatial and a quasi-Gaussian decay of temporal correlations~\cite{de_giorgi_interaction_2018}. In polariton condensates, temporal correlation decay has been reported to be Gaussian~\cite{baboux_unstable_2018,love_intrinsic_2008,kim_coherent_2016,daskalakis_spatial_2015,haug_temporal_2012} or of Kubo form, i.e. Gaussian or exponential depending on whether the number fluctuations are slow or fast~\cite{whittaker_coherence_2009,spano_build_2013}.

	Above the BEC threshold, both spatial and temporal correlations show decay that indicates power-law behaviour. Spatial correlations decay with a small exponent of $0.07$ whereas temporal correlations with a large exponent of around $0.7-0.8$. This clearly differs from equilibrium BKT transition, as could be expected due to the driven-dissipative nature of our system. It is worthwhile to mention that we have also looked at single-shot interferograms of the condensate and have not observed vortices. In driven-dissipative condensates, the power-law exponent may exceed the equilibrium BKT limit $0.25$, but our findings do not directly match with the reported experimental~\cite{roumpos_power-law_2012} or theoretical values~\cite{dagvadorj_nonequilibrium_2015,comaron_non-equilibrium_2021}.

	Stretched exponential function, related to KPZ dynamics, fits approximately equally well as the power law, yielding small exponents $\beta$. The KPZ scaling of correlations has been mostly considered in the optical parametric oscillator regime of microcavity polariton condensates~\cite{zamora_tuning_2017,ferrier_searching_2020}. In the strongly anisotropic case, the KPZ equation leads to a power-law decay of spatial correlations with exponent $b_\mathrm{s}=2\chi$, where $\chi$ is the roughness exponent that in 2D takes the universal value of $0.39$~\cite{halpin-healy_universal_2014,pagnani_numerical_2015,miranda_numerical_2008}. Likewise, the temporal correlations decay as power law but with exponent that is 1/2 of the spatial exponent, $b_\mathrm{t}=\chi$. In the regime of weak anisotropy, the 2D KPZ equation predicts a stretched exponential decay with $\beta_\mathrm{s}=2\chi=0.78$ for spatial and $\beta_\mathrm{t}=2\chi/(2-\chi)=0.48$ for temporal correlations~\cite{zamora_tuning_2017}. Our results yield $\beta_\mathrm{s} \approx 0.1$ and $\beta_\mathrm{t} \approx 0.2$ above the BEC threshold.
	
	Our system is not isotropic. The nanoparticles host dipolar and multipolar charge oscillation modes associated with directional radiation, which combined with the periodicity of the array in the $x$ and $y$ directions leads to SLR modes that are not isotropic in the plane. In an infinite or square array, there is still $x$-$y$ symmetry, but even this can be broken by the pump polarization (see \cite{SM} for further discussion). Evaluating the level of anisotropy in the sense of the theoretical predictions is difficult since our system is not analogous to those considered in Refs.~\cite{zamora_tuning_2017,altman_two-dimensional_2015}, however the KPZ scaling is expected to be rather universal.
	
	Finally, it should be noted that the differences (e.g. RMSE) between fits 
	to different models have typically been marginal, and therefore 
	establishing power-law behaviour quantitatively is not straightforward
	~\cite{caputo_topological_2018,roumpos_power-law_2012,nitsche_algebraic_2014,zamora_tuning_2017}. 
	Ideally the data should range over at least two decades to rigorously claim power-law behaviour~\cite{clauset_power-law_2009}. The results depend on the fit range, so it is of key importance to justify the range (short- vs. long-range) deliberately. Another remark is that the regime of pump powers where the transition to a BKT phase or KPZ dynamics is expected could be very narrow~\cite{comaron_dynamical_2018,comaron_non-equilibrium_2021,dagvadorj_nonequilibrium_2015}.
	
	\textit{Summary and outlook.}---We have characterized for the first time the spatial and temporal first-order correlations in a plasmonic BEC. We propose plasmonic lattice condensates as a platform for studying long-range order in 2D condensates of light, with versatile system design and open-cavity character. The combination of large size and short cavity lifetime makes the system amenable for probing non-equilibrium BKT transition or KPZ dynamics~\cite{he_scaling_2015,altman_two-dimensional_2015,comaron_dynamical_2018,zamora_tuning_2017}. 
	
	In this work, we have shown that the decay of correlations above the BEC threshold is non-exponential, in stark contrast to the non-ordered and lasing phases. Our results indicate algebraic decay of both spatial and temporal correlations with different exponents, however the stretched exponential fits approximately equally well. The exponents found do not quantitatively match with reported predictions for equilibrium or non-equilibrium systems. Although the scaling laws predicted by the equilibrium BKT and the non-equilibrium KPZ mechanisms might be universal, understanding the role of strongly coupled
	photonic, electronic, and vibrational states in the condensate formation is an important general question that requires future theory work and may bring in new physics. Moreover, the known plasmonic nanoparticle array mode structure~\cite{Moerland2017,guo_geometry_2017,wang_rich_2018,kravets_plasmonic_2018,wang_structural_2018} should be incorporated in the theoretical analysis to quantify the degree of anisotropy. 
	
	Beyond fundamental studies of correlations, the plasmonic polariton BEC provides spatial coherence decay on the millimeter-scale, which is 1--2 orders of magnitude larger than in other polariton or photon condensates before. Millimeter-scale spatial coherence has been previously reported in weakly coupled plasmonic lasers~\cite{hoang_millimeter-scale_2017}. Strongly coupled plasmonic BECs provide also effective interactions~\cite{Vakevainen2020}, which may give rise to phenomena not accessible in weakly coupled and non-interacting systems, such as superfluidity~\cite{keeling_superfluidity_2017}. The large extent of spatial coherence could be utilized in, for instance, on-chip applications for sensing, where lasing and condensation can make tiny effects observable.
	
	{\em Acknowledgments---}
	We thank Jonathan Keeling and Kristin B. Arnardottir for their comments on the early version of the manuscript. AJM and PT acknowledge support by the Academy of Finland under project numbers 303351, 327293, 318937
	(PROFI), and 320167 (Flagship Programme Photonics Research and Innovation (PREIN)). AJM acknowledges financial support by the Jenny and Antti Wihuri Foundation. KD acknowledges financial support from the European Research Council (ERC) under the European Union’s Horizon 2020 research and innovation programme (grant agreement No. [948260]).

	\bibliographystyle{apsrev4-1_abbrv}
	\bibliography{References}

\begin{thebibliography}{57}%
\makeatletter
\providecommand \@ifxundefined [1]{%
 \@ifx{#1\undefined}
}%
\providecommand \@ifnum [1]{%
 \ifnum #1\expandafter \@firstoftwo
 \else \expandafter \@secondoftwo
 \fi
}%
\providecommand \@ifx [1]{%
 \ifx #1\expandafter \@firstoftwo
 \else \expandafter \@secondoftwo
 \fi
}%
\providecommand \natexlab [1]{#1}%
\providecommand \enquote  [1]{``#1''}%
\providecommand \bibnamefont  [1]{#1}%
\providecommand \bibfnamefont [1]{#1}%
\providecommand \citenamefont [1]{#1}%
\providecommand \href@noop [0]{\@secondoftwo}%
\providecommand \href [0]{\begingroup \@sanitize@url \@href}%
\providecommand \@href[1]{\@@startlink{#1}\@@href}%
\providecommand \@@href[1]{\endgroup#1\@@endlink}%
\providecommand \@sanitize@url [0]{\catcode `\\12\catcode `\$12\catcode
  `\&12\catcode `\#12\catcode `\^12\catcode `\_12\catcode `\%12\relax}%
\providecommand \@@startlink[1]{}%
\providecommand \@@endlink[0]{}%
\providecommand \url  [0]{\begingroup\@sanitize@url \@url }%
\providecommand \@url [1]{\endgroup\@href {#1}{\urlprefix }}%
\providecommand \urlprefix  [0]{URL }%
\providecommand \Eprint [0]{\href }%
\providecommand \doibase [0]{http://dx.doi.org/}%
\providecommand \selectlanguage [0]{\@gobble}%
\providecommand \bibinfo  [0]{\@secondoftwo}%
\providecommand \bibfield  [0]{\@secondoftwo}%
\providecommand \translation [1]{[#1]}%
\providecommand \BibitemOpen [0]{}%
\providecommand \bibitemStop [0]{}%
\providecommand \bibitemNoStop [0]{.\EOS\space}%
\providecommand \EOS [0]{\spacefactor3000\relax}%
\providecommand \BibitemShut  [1]{\csname bibitem#1\endcsname}%
\let\auto@bib@innerbib\@empty
\bibitem [{\citenamefont {Mermin}\ and\ \citenamefont
  {Wagner}(1966)}]{mermin_absence_1966}%
  \BibitemOpen
  \bibfield  {author} {\bibinfo {author} {\bibfnamefont {N.~D.}\ \bibnamefont
  {Mermin}}\ and\ \bibinfo {author} {\bibfnamefont {H.}~\bibnamefont
  {Wagner}},\ }\bibfield  {title} {\bibinfo {title} {\emph {Absence of
  {Ferromagnetism} or {Antiferromagnetism} in {One}- or {Two}-{Dimensional}
  {Isotropic} {Heisenberg} {Models}}},\ }\href {\doibase
  10.1103/PhysRevLett.17.1133} {\bibfield  {journal} {\bibinfo  {journal}
  {Physical Review Letters}\ }\textbf {\bibinfo {volume} {17}},\ \bibinfo
  {pages} {1133} (\bibinfo {year} {1966})}\BibitemShut {NoStop}%
\bibitem [{\citenamefont {Hohenberg}(1967)}]{hohenberg_existence_1967}%
  \BibitemOpen
  \bibfield  {author} {\bibinfo {author} {\bibfnamefont {P.~C.}\ \bibnamefont
  {Hohenberg}},\ }\bibfield  {title} {\bibinfo {title} {\emph {Existence of
  {Long}-{Range} {Order} in {One} and {Two} {Dimensions}}},\ }\href {\doibase
  10.1103/PhysRev.158.383} {\bibfield  {journal} {\bibinfo  {journal} {Physical
  Review}\ }\textbf {\bibinfo {volume} {158}},\ \bibinfo {pages} {383}
  (\bibinfo {year} {1967})}\BibitemShut {NoStop}%
\bibitem [{\citenamefont {Bloch}\ \emph {et~al.}(2008)\citenamefont {Bloch},
  \citenamefont {Dalibard},\ and\ \citenamefont
  {Zwerger}}]{bloch_many-body_2008}%
  \BibitemOpen
  \bibfield  {author} {\bibinfo {author} {\bibfnamefont {I.}~\bibnamefont
  {Bloch}}, \bibinfo {author} {\bibfnamefont {J.}~\bibnamefont {Dalibard}}, \
  and\ \bibinfo {author} {\bibfnamefont {W.}~\bibnamefont {Zwerger}},\
  }\bibfield  {title} {\bibinfo {title} {\emph {Many-Body Physics with
  Ultracold Gases}},\ }\href {\doibase 10.1103/RevModPhys.80.885} {\bibfield
  {journal} {\bibinfo  {journal} {Reviews of Modern Physics}\ }\textbf
  {\bibinfo {volume} {80}},\ \bibinfo {pages} {885} (\bibinfo {year}
  {2008})}\BibitemShut {NoStop}%
\bibitem [{\citenamefont {Kosterlitz}\ and\ \citenamefont
  {Thouless}(1973)}]{kosterlitz_ordering_1973}%
  \BibitemOpen
  \bibfield  {author} {\bibinfo {author} {\bibfnamefont {J.~M.}\ \bibnamefont
  {Kosterlitz}}\ and\ \bibinfo {author} {\bibfnamefont {D.~J.}\ \bibnamefont
  {Thouless}},\ }\bibfield  {title} {\bibinfo {title} {\emph {Ordering,
  metastability and phase transitions in two-dimensional systems}},\ }\href
  {\doibase 10.1088/0022-3719/6/7/010} {\bibfield  {journal} {\bibinfo
  {journal} {Journal of Physics C: Solid State Physics}\ }\textbf {\bibinfo
  {volume} {6}},\ \bibinfo {pages} {1181} (\bibinfo {year} {1973})}\BibitemShut
  {NoStop}%
\bibitem [{\citenamefont {Berezinskii}(1971)}]{berezinskii_1971}%
  \BibitemOpen
  \bibfield  {author} {\bibinfo {author} {\bibfnamefont {V.~L.}\ \bibnamefont
  {Berezinskii}},\ }\bibfield  {title} {\bibinfo {title} {\emph {Destruction of
  Long-range Order in One-dimensional and Two-dimensional Systems having a
  Continuous Symmetry Group I. Classical Systems}},\ }\href@noop {} {\bibfield
  {journal} {\bibinfo  {journal} {Soviet Physics JETP}\ }\textbf {\bibinfo
  {volume} {32}},\ \bibinfo {pages} {493} (\bibinfo {year} {1971})}\BibitemShut
  {NoStop}%
\bibitem [{\citenamefont {Kardar}\ \emph {et~al.}(1986)\citenamefont {Kardar},
  \citenamefont {Parisi},\ and\ \citenamefont {Zhang}}]{kardar_dynamic_1986}%
  \BibitemOpen
  \bibfield  {author} {\bibinfo {author} {\bibfnamefont {M.}~\bibnamefont
  {Kardar}}, \bibinfo {author} {\bibfnamefont {G.}~\bibnamefont {Parisi}}, \
  and\ \bibinfo {author} {\bibfnamefont {Y.-C.}\ \bibnamefont {Zhang}},\
  }\bibfield  {title} {\bibinfo {title} {\emph {Dynamic {Scaling} of {Growing}
  {Interfaces}}},\ }\href {\doibase 10.1103/PhysRevLett.56.889} {\bibfield
  {journal} {\bibinfo  {journal} {Physical Review Letters}\ }\textbf {\bibinfo
  {volume} {56}},\ \bibinfo {pages} {889} (\bibinfo {year} {1986})}\BibitemShut
  {NoStop}%
\bibitem [{\citenamefont {Caputo}\ \emph {et~al.}(2018)\citenamefont {Caputo},
  \citenamefont {Ballarini}, \citenamefont {Dagvadorj}, \citenamefont
  {Sánchez Muñoz}, \citenamefont {De Giorgi}, \citenamefont {Dominici},
  \citenamefont {West}, \citenamefont {Pfeiffer}, \citenamefont {Gigli},
  \citenamefont {Laussy}, \citenamefont {Szymanska},\ and\ \citenamefont
  {Sanvitto}}]{caputo_topological_2018}%
  \BibitemOpen
  \bibfield  {author} {\bibinfo {author} {\bibfnamefont {D.}~\bibnamefont
  {Caputo}}, \bibinfo {author} {\bibfnamefont {D.}~\bibnamefont {Ballarini}},
  \bibinfo {author} {\bibfnamefont {G.}~\bibnamefont {Dagvadorj}}, \bibinfo
  {author} {\bibfnamefont {C.}~\bibnamefont {Sánchez Muñoz}}, \bibinfo
  {author} {\bibfnamefont {M.}~\bibnamefont {De Giorgi}}, \bibinfo {author}
  {\bibfnamefont {L.}~\bibnamefont {Dominici}}, \bibinfo {author}
  {\bibfnamefont {K.}~\bibnamefont {West}}, \bibinfo {author} {\bibfnamefont
  {L.~N.}\ \bibnamefont {Pfeiffer}}, \bibinfo {author} {\bibfnamefont
  {G.}~\bibnamefont {Gigli}}, \bibinfo {author} {\bibfnamefont {F.~P.}\
  \bibnamefont {Laussy}}, \bibinfo {author} {\bibfnamefont {M.~H.}\
  \bibnamefont {Szymanska}}, \ and\ \bibinfo {author} {\bibfnamefont
  {D.}~\bibnamefont {Sanvitto}},\ }\bibfield  {title} {\bibinfo {title} {\emph
  {Topological order and thermal equilibrium in polariton condensates}},\
  }\href {\doibase 10.1038/nmat5039} {\bibfield  {journal} {\bibinfo  {journal}
  {Nature Materials}\ }\textbf {\bibinfo {volume} {17}},\ \bibinfo {pages}
  {145} (\bibinfo {year} {2018})}\BibitemShut {NoStop}%
\bibitem [{\citenamefont {Daskalakis}\ \emph {et~al.}(2015)\citenamefont
  {Daskalakis}, \citenamefont {Maier},\ and\ \citenamefont
  {Kena-Cohen}}]{daskalakis_spatial_2015}%
  \BibitemOpen
  \bibfield  {author} {\bibinfo {author} {\bibfnamefont {K.~S.}\ \bibnamefont
  {Daskalakis}}, \bibinfo {author} {\bibfnamefont {S.~A.}\ \bibnamefont
  {Maier}}, \ and\ \bibinfo {author} {\bibfnamefont {S.}~\bibnamefont
  {Kena-Cohen}},\ }\bibfield  {title} {\bibinfo {title} {\emph {Spatial
  {Coherence} and {Stability} in a {Disordered} {Organic} {Polariton}
  {Condensate}}},\ }\href {\doibase 10.1103/PhysRevLett.115.035301} {\bibfield
  {journal} {\bibinfo  {journal} {Physical Review Letters}\ }\textbf {\bibinfo
  {volume} {115}},\ \bibinfo {pages} {035301} (\bibinfo {year}
  {2015})}\BibitemShut {NoStop}%
\bibitem [{\citenamefont {Damm}\ \emph {et~al.}(2017)\citenamefont {Damm},
  \citenamefont {Dung}, \citenamefont {Vewinger}, \citenamefont {Weitz},\ and\
  \citenamefont {Schmitt}}]{Damm2017NatComm}%
  \BibitemOpen
  \bibfield  {author} {\bibinfo {author} {\bibfnamefont {T.}~\bibnamefont
  {Damm}}, \bibinfo {author} {\bibfnamefont {D.}~\bibnamefont {Dung}}, \bibinfo
  {author} {\bibfnamefont {F.}~\bibnamefont {Vewinger}}, \bibinfo {author}
  {\bibfnamefont {M.}~\bibnamefont {Weitz}}, \ and\ \bibinfo {author}
  {\bibfnamefont {J.}~\bibnamefont {Schmitt}},\ }\bibfield  {title} {\bibinfo
  {title} {\emph {First-order spatial coherence measurements in a thermalized
  two-dimensional photonic quantum gas}},\ }\href {\doibase
  10.1038/s41467-017-00270-8} {\bibfield  {journal} {\bibinfo  {journal}
  {Nature Communications}\ }\textbf {\bibinfo {volume} {8}},\ \bibinfo {pages}
  {158} (\bibinfo {year} {2017})}\BibitemShut {NoStop}%
\bibitem [{\citenamefont {Marelic}\ \emph {et~al.}(2016)\citenamefont
  {Marelic}, \citenamefont {Zajiczek}, \citenamefont {Hesten}, \citenamefont
  {Leung}, \citenamefont {Ong}, \citenamefont {Mintert},\ and\ \citenamefont
  {Nyman}}]{marelic_spatiotemporal_2016}%
  \BibitemOpen
  \bibfield  {author} {\bibinfo {author} {\bibfnamefont {J.}~\bibnamefont
  {Marelic}}, \bibinfo {author} {\bibfnamefont {L.~F.}\ \bibnamefont
  {Zajiczek}}, \bibinfo {author} {\bibfnamefont {H.~J.}\ \bibnamefont
  {Hesten}}, \bibinfo {author} {\bibfnamefont {K.~H.}\ \bibnamefont {Leung}},
  \bibinfo {author} {\bibfnamefont {E.~Y.~X.}\ \bibnamefont {Ong}}, \bibinfo
  {author} {\bibfnamefont {F.}~\bibnamefont {Mintert}}, \ and\ \bibinfo
  {author} {\bibfnamefont {R.~A.}\ \bibnamefont {Nyman}},\ }\bibfield  {title}
  {\bibinfo {title} {\emph {Spatiotemporal coherence of non-equilibrium
  multimode photon condensates}},\ }\href {\doibase
  10.1088/1367-2630/18/10/103012} {\bibfield  {journal} {\bibinfo  {journal}
  {New Journal of Physics}\ }\textbf {\bibinfo {volume} {18}},\ \bibinfo
  {pages} {103012} (\bibinfo {year} {2016})}\BibitemShut {NoStop}%
\bibitem [{\citenamefont {Nelson}\ and\ \citenamefont
  {Kosterlitz}(1977)}]{nelson_universal_1977}%
  \BibitemOpen
  \bibfield  {author} {\bibinfo {author} {\bibfnamefont {D.~R.}\ \bibnamefont
  {Nelson}}\ and\ \bibinfo {author} {\bibfnamefont {J.~M.}\ \bibnamefont
  {Kosterlitz}},\ }\bibfield  {title} {\bibinfo {title} {\emph {Universal
  {Jump} in the {Superfluid} {Density} of {Two}-{Dimensional} {Superfluids}}},\
  }\href {\doibase 10.1103/PhysRevLett.39.1201} {\bibfield  {journal} {\bibinfo
   {journal} {Physical Review Letters}\ }\textbf {\bibinfo {volume} {39}},\
  \bibinfo {pages} {1201} (\bibinfo {year} {1977})}\BibitemShut {NoStop}%
\bibitem [{\citenamefont {Comaron}\ \emph {et~al.}(2021)\citenamefont
  {Comaron}, \citenamefont {Carusotto}, \citenamefont {Szymanska},\ and\
  \citenamefont {Proukakis}}]{comaron_non-equilibrium_2021}%
  \BibitemOpen
  \bibfield  {author} {\bibinfo {author} {\bibfnamefont {P.}~\bibnamefont
  {Comaron}}, \bibinfo {author} {\bibfnamefont {I.}~\bibnamefont {Carusotto}},
  \bibinfo {author} {\bibfnamefont {M.~H.}\ \bibnamefont {Szymanska}}, \ and\
  \bibinfo {author} {\bibfnamefont {N.~P.}\ \bibnamefont {Proukakis}},\
  }\bibfield  {title} {\bibinfo {title} {\emph {Non-equilibrium
  {Berezinskii}-{Kosterlitz}-{Thouless} transition in driven-dissipative
  condensates (a)}},\ }\href {\doibase 10.1209/0295-5075/133/17002} {\bibfield
  {journal} {\bibinfo  {journal} {EPL (Europhysics Letters)}\ }\textbf
  {\bibinfo {volume} {133}},\ \bibinfo {pages} {17002} (\bibinfo {year}
  {2021})}\BibitemShut {NoStop}%
\bibitem [{\citenamefont {Szymanska}\ \emph {et~al.}(2007)\citenamefont
  {Szymanska}, \citenamefont {Keeling},\ and\ \citenamefont
  {Littlewood}}]{szymanska_mean-field_2007}%
  \BibitemOpen
  \bibfield  {author} {\bibinfo {author} {\bibfnamefont {M.~H.}\ \bibnamefont
  {Szymanska}}, \bibinfo {author} {\bibfnamefont {J.}~\bibnamefont {Keeling}},
  \ and\ \bibinfo {author} {\bibfnamefont {P.~B.}\ \bibnamefont {Littlewood}},\
  }\bibfield  {title} {\bibinfo {title} {\emph {Mean-field theory and
  fluctuation spectrum of a pumped decaying {Bose}-{Fermi} system across the
  quantum condensation transition}},\ }\href {\doibase
  10.1103/PhysRevB.75.195331} {\bibfield  {journal} {\bibinfo  {journal}
  {Physical Review B}\ }\textbf {\bibinfo {volume} {75}},\ \bibinfo {pages}
  {195331} (\bibinfo {year} {2007})}\BibitemShut {NoStop}%
\bibitem [{\citenamefont {Dagvadorj}\ \emph {et~al.}(2015)\citenamefont
  {Dagvadorj}, \citenamefont {Fellows}, \citenamefont {Matyjaśkiewicz},
  \citenamefont {Marchetti}, \citenamefont {Carusotto},\ and\ \citenamefont
  {Szymanska}}]{dagvadorj_nonequilibrium_2015}%
  \BibitemOpen
  \bibfield  {author} {\bibinfo {author} {\bibfnamefont {G.}~\bibnamefont
  {Dagvadorj}}, \bibinfo {author} {\bibfnamefont {J.~M.}\ \bibnamefont
  {Fellows}}, \bibinfo {author} {\bibfnamefont {S.}~\bibnamefont
  {Matyjaśkiewicz}}, \bibinfo {author} {\bibfnamefont {F.~M.}\ \bibnamefont
  {Marchetti}}, \bibinfo {author} {\bibfnamefont {I.}~\bibnamefont
  {Carusotto}}, \ and\ \bibinfo {author} {\bibfnamefont {M.~H.}\ \bibnamefont
  {Szymanska}},\ }\bibfield  {title} {\bibinfo {title} {\emph {Nonequilibrium
  {Phase} {Transition} in a {Two}-{Dimensional} {Driven} {Open} {Quantum}
  {System}}},\ }\href {\doibase 10.1103/PhysRevX.5.041028} {\bibfield
  {journal} {\bibinfo  {journal} {Physical Review X}\ }\textbf {\bibinfo
  {volume} {5}},\ \bibinfo {pages} {041028} (\bibinfo {year}
  {2015})}\BibitemShut {NoStop}%
\bibitem [{\citenamefont {Altman}\ \emph {et~al.}(2015)\citenamefont {Altman},
  \citenamefont {Sieberer}, \citenamefont {Chen}, \citenamefont {Diehl},\ and\
  \citenamefont {Toner}}]{altman_two-dimensional_2015}%
  \BibitemOpen
  \bibfield  {author} {\bibinfo {author} {\bibfnamefont {E.}~\bibnamefont
  {Altman}}, \bibinfo {author} {\bibfnamefont {L.~M.}\ \bibnamefont
  {Sieberer}}, \bibinfo {author} {\bibfnamefont {L.}~\bibnamefont {Chen}},
  \bibinfo {author} {\bibfnamefont {S.}~\bibnamefont {Diehl}}, \ and\ \bibinfo
  {author} {\bibfnamefont {J.}~\bibnamefont {Toner}},\ }\bibfield  {title}
  {\bibinfo {title} {\emph {Two-{{Dimensional Superfluidity}} of {{Exciton
  Polaritons Requires Strong Anisotropy}}}},\ }\href {\doibase
  10.1103/PhysRevX.5.011017} {\bibfield  {journal} {\bibinfo  {journal}
  {Physical Review X}\ }\textbf {\bibinfo {volume} {5}},\ \bibinfo {pages}
  {011017} (\bibinfo {year} {2015})}\BibitemShut {NoStop}%
\bibitem [{\citenamefont {Ferrier}\ \emph {et~al.}(2020)\citenamefont
  {Ferrier}, \citenamefont {Zamora}, \citenamefont {Dagvadorj},\ and\
  \citenamefont {Szymanska}}]{ferrier_searching_2020}%
  \BibitemOpen
  \bibfield  {author} {\bibinfo {author} {\bibfnamefont {A.}~\bibnamefont
  {Ferrier}}, \bibinfo {author} {\bibfnamefont {A.}~\bibnamefont {Zamora}},
  \bibinfo {author} {\bibfnamefont {G.}~\bibnamefont {Dagvadorj}}, \ and\
  \bibinfo {author} {\bibfnamefont {M.~H.}\ \bibnamefont {Szymanska}},\
  }\bibfield  {title} {\bibinfo {title} {\emph {Searching for the
  {Kardar}-{Parisi}-{Zhang} phase in microcavity polaritons}},\ }\href
  {http://arxiv.org/abs/2009.05177} {\bibfield  {journal} {\bibinfo  {journal}
  {arXiv:2009.05177 [cond-mat]}\ } (\bibinfo {year} {2020})}\BibitemShut
  {NoStop}%
\bibitem [{\citenamefont {Comaron}\ \emph {et~al.}(2018)\citenamefont
  {Comaron}, \citenamefont {Dagvadorj}, \citenamefont {Zamora}, \citenamefont
  {Carusotto}, \citenamefont {Proukakis},\ and\ \citenamefont
  {Szymanska}}]{comaron_dynamical_2018}%
  \BibitemOpen
  \bibfield  {author} {\bibinfo {author} {\bibfnamefont {P.}~\bibnamefont
  {Comaron}}, \bibinfo {author} {\bibfnamefont {G.}~\bibnamefont {Dagvadorj}},
  \bibinfo {author} {\bibfnamefont {A.}~\bibnamefont {Zamora}}, \bibinfo
  {author} {\bibfnamefont {I.}~\bibnamefont {Carusotto}}, \bibinfo {author}
  {\bibfnamefont {N.~P.}\ \bibnamefont {Proukakis}}, \ and\ \bibinfo {author}
  {\bibfnamefont {M.~H.}\ \bibnamefont {Szymanska}},\ }\bibfield  {title}
  {\bibinfo {title} {\emph {Dynamical {Critical} {Exponents} in
  {Driven}-{Dissipative} {Quantum} {Systems}}},\ }\href {\doibase
  10.1103/PhysRevLett.121.095302} {\bibfield  {journal} {\bibinfo  {journal}
  {Physical Review Letters}\ }\textbf {\bibinfo {volume} {121}},\ \bibinfo
  {pages} {095302} (\bibinfo {year} {2018})}\BibitemShut {NoStop}%
\bibitem [{\citenamefont {Zamora}\ \emph {et~al.}(2017)\citenamefont {Zamora},
  \citenamefont {Sieberer}, \citenamefont {Dunnett}, \citenamefont {Diehl},\
  and\ \citenamefont {Szymanska}}]{zamora_tuning_2017}%
  \BibitemOpen
  \bibfield  {author} {\bibinfo {author} {\bibfnamefont {A.}~\bibnamefont
  {Zamora}}, \bibinfo {author} {\bibfnamefont {L.~M.}\ \bibnamefont
  {Sieberer}}, \bibinfo {author} {\bibfnamefont {K.}~\bibnamefont {Dunnett}},
  \bibinfo {author} {\bibfnamefont {S.}~\bibnamefont {Diehl}}, \ and\ \bibinfo
  {author} {\bibfnamefont {M.~H.}\ \bibnamefont {Szymanska}},\ }\bibfield
  {title} {\bibinfo {title} {\emph {Tuning across {Universalities} with a
  {Driven} {Open} {Condensate}}},\ }\href {\doibase 10.1103/PhysRevX.7.041006}
  {\bibfield  {journal} {\bibinfo  {journal} {Physical Review X}\ }\textbf
  {\bibinfo {volume} {7}},\ \bibinfo {pages} {041006} (\bibinfo {year}
  {2017})}\BibitemShut {NoStop}%
\bibitem [{\citenamefont {Schawlow}\ and\ \citenamefont
  {Townes}(1958)}]{schawlow_infrared_1958}%
  \BibitemOpen
  \bibfield  {author} {\bibinfo {author} {\bibfnamefont {A.~L.}\ \bibnamefont
  {Schawlow}}\ and\ \bibinfo {author} {\bibfnamefont {C.~H.}\ \bibnamefont
  {Townes}},\ }\bibfield  {title} {\bibinfo {title} {\emph {Infrared and
  {Optical} {Masers}}},\ }\href {\doibase 10.1103/PhysRev.112.1940} {\bibfield
  {journal} {\bibinfo  {journal} {Physical Review}\ }\textbf {\bibinfo {volume}
  {112}},\ \bibinfo {pages} {1940} (\bibinfo {year} {1958})}\BibitemShut
  {NoStop}%
\bibitem [{\citenamefont {Kasprzak}\ \emph {et~al.}(2006)\citenamefont
  {Kasprzak}, \citenamefont {Richard}, \citenamefont {Kundermann},
  \citenamefont {Baas}, \citenamefont {Jeambrun}, \citenamefont {Keeling},
  \citenamefont {Marchetti}, \citenamefont {Szymanska}, \citenamefont {Andre},
  \citenamefont {Staehli}, \citenamefont {Savona}, \citenamefont {Littlewood},
  \citenamefont {Deveaud},\ and\ \citenamefont
  {Dang}}]{kasprzak_bose-einstein_2006}%
  \BibitemOpen
  \bibfield  {author} {\bibinfo {author} {\bibfnamefont {J.}~\bibnamefont
  {Kasprzak}}, \bibinfo {author} {\bibfnamefont {M.}~\bibnamefont {Richard}},
  \bibinfo {author} {\bibfnamefont {S.}~\bibnamefont {Kundermann}}, \bibinfo
  {author} {\bibfnamefont {A.}~\bibnamefont {Baas}}, \bibinfo {author}
  {\bibfnamefont {P.}~\bibnamefont {Jeambrun}}, \bibinfo {author}
  {\bibfnamefont {J.~M.~J.}\ \bibnamefont {Keeling}}, \bibinfo {author}
  {\bibfnamefont {F.~M.}\ \bibnamefont {Marchetti}}, \bibinfo {author}
  {\bibfnamefont {M.~H.}\ \bibnamefont {Szymanska}}, \bibinfo {author}
  {\bibfnamefont {R.}~\bibnamefont {Andre}}, \bibinfo {author} {\bibfnamefont
  {J.~L.}\ \bibnamefont {Staehli}}, \bibinfo {author} {\bibfnamefont
  {V.}~\bibnamefont {Savona}}, \bibinfo {author} {\bibfnamefont {P.~B.}\
  \bibnamefont {Littlewood}}, \bibinfo {author} {\bibfnamefont
  {B.}~\bibnamefont {Deveaud}}, \ and\ \bibinfo {author} {\bibfnamefont
  {L.~S.}\ \bibnamefont {Dang}},\ }\bibfield  {title} {\bibinfo {title} {\emph
  {Bose\textendash{{Einstein}} Condensation of Exciton Polaritons}},\ }\href
  {\doibase 10.1038/nature05131} {\bibfield  {journal} {\bibinfo  {journal}
  {Nature}\ }\textbf {\bibinfo {volume} {443}},\ \bibinfo {pages} {409}
  (\bibinfo {year} {2006})}\BibitemShut {NoStop}%
\bibitem [{\citenamefont {Daskalakis}\ \emph {et~al.}(2014)\citenamefont
  {Daskalakis}, \citenamefont {Maier}, \citenamefont {Murray},\ and\
  \citenamefont {{K{\'e}na-Cohen}}}]{daskalakis_nonlinear_2014}%
  \BibitemOpen
  \bibfield  {author} {\bibinfo {author} {\bibfnamefont {K.~S.}\ \bibnamefont
  {Daskalakis}}, \bibinfo {author} {\bibfnamefont {S.~A.}\ \bibnamefont
  {Maier}}, \bibinfo {author} {\bibfnamefont {R.}~\bibnamefont {Murray}}, \
  and\ \bibinfo {author} {\bibfnamefont {S.}~\bibnamefont {{K{\'e}na-Cohen}}},\
  }\bibfield  {title} {\bibinfo {title} {\emph {Nonlinear Interactions in an
  Organic Polariton Condensate}},\ }\href {\doibase 10.1038/nmat3874}
  {\bibfield  {journal} {\bibinfo  {journal} {Nature Materials}\ }\textbf
  {\bibinfo {volume} {13}},\ \bibinfo {pages} {271} (\bibinfo {year}
  {2014})}\BibitemShut {NoStop}%
\bibitem [{\citenamefont {Plumhof}\ \emph {et~al.}(2014)\citenamefont
  {Plumhof}, \citenamefont {St{\"o}ferle}, \citenamefont {Mai}, \citenamefont
  {Scherf},\ and\ \citenamefont {Mahrt}}]{plumhof_room-temperature_2014}%
  \BibitemOpen
  \bibfield  {author} {\bibinfo {author} {\bibfnamefont {J.~D.}\ \bibnamefont
  {Plumhof}}, \bibinfo {author} {\bibfnamefont {T.}~\bibnamefont
  {St{\"o}ferle}}, \bibinfo {author} {\bibfnamefont {L.}~\bibnamefont {Mai}},
  \bibinfo {author} {\bibfnamefont {U.}~\bibnamefont {Scherf}}, \ and\ \bibinfo
  {author} {\bibfnamefont {R.~F.}\ \bibnamefont {Mahrt}},\ }\bibfield  {title}
  {\bibinfo {title} {\emph {Room-Temperature {{Bose}}\textendash{{Einstein}}
  Condensation of Cavity Exciton\textendash{}Polaritons in a Polymer}},\ }\href
  {\doibase 10.1038/nmat3825} {\bibfield  {journal} {\bibinfo  {journal}
  {Nature Materials}\ }\textbf {\bibinfo {volume} {13}},\ \bibinfo {pages}
  {247} (\bibinfo {year} {2014})}\BibitemShut {NoStop}%
\bibitem [{\citenamefont {Carusotto}\ and\ \citenamefont
  {Ciuti}(2013)}]{carusotto_quantum_2013}%
  \BibitemOpen
  \bibfield  {author} {\bibinfo {author} {\bibfnamefont {I.}~\bibnamefont
  {Carusotto}}\ and\ \bibinfo {author} {\bibfnamefont {C.}~\bibnamefont
  {Ciuti}},\ }\bibfield  {title} {\bibinfo {title} {\emph {Quantum Fluids of
  Light}},\ }\href {\doibase 10.1103/RevModPhys.85.299} {\bibfield  {journal}
  {\bibinfo  {journal} {Reviews of Modern Physics}\ }\textbf {\bibinfo {volume}
  {85}},\ \bibinfo {pages} {299} (\bibinfo {year} {2013})}\BibitemShut
  {NoStop}%
\bibitem [{\citenamefont {Keeling}\ and\ \citenamefont
  {Kena-Cohen}(2020)}]{keeling_boseeinstein_2020}%
  \BibitemOpen
  \bibfield  {author} {\bibinfo {author} {\bibfnamefont {J.}~\bibnamefont
  {Keeling}}\ and\ \bibinfo {author} {\bibfnamefont {S.}~\bibnamefont
  {Kena-Cohen}},\ }\bibfield  {title} {\bibinfo {title} {\emph
  {Bose–{Einstein} {Condensation} of {Exciton}-{Polaritons} in {Organic}
  {Microcavities}}},\ }\href {\doibase 10.1146/annurev-physchem-010920-102509}
  {\bibfield  {journal} {\bibinfo  {journal} {Annual Review of Physical
  Chemistry}\ }\textbf {\bibinfo {volume} {71}},\ \bibinfo {pages} {435}
  (\bibinfo {year} {2020})}\BibitemShut {NoStop}%
\bibitem [{\citenamefont {Sun}\ \emph {et~al.}(2017)\citenamefont {Sun},
  \citenamefont {Wen}, \citenamefont {Yoon}, \citenamefont {Liu}, \citenamefont
  {Steger}, \citenamefont {Pfeiffer}, \citenamefont {West}, \citenamefont
  {Snoke},\ and\ \citenamefont {Nelson}}]{sun_bose-einstein_2017}%
  \BibitemOpen
  \bibfield  {author} {\bibinfo {author} {\bibfnamefont {Y.}~\bibnamefont
  {Sun}}, \bibinfo {author} {\bibfnamefont {P.}~\bibnamefont {Wen}}, \bibinfo
  {author} {\bibfnamefont {Y.}~\bibnamefont {Yoon}}, \bibinfo {author}
  {\bibfnamefont {G.}~\bibnamefont {Liu}}, \bibinfo {author} {\bibfnamefont
  {M.}~\bibnamefont {Steger}}, \bibinfo {author} {\bibfnamefont {L.~N.}\
  \bibnamefont {Pfeiffer}}, \bibinfo {author} {\bibfnamefont {K.}~\bibnamefont
  {West}}, \bibinfo {author} {\bibfnamefont {D.~W.}\ \bibnamefont {Snoke}}, \
  and\ \bibinfo {author} {\bibfnamefont {K.~A.}\ \bibnamefont {Nelson}},\
  }\bibfield  {title} {\bibinfo {title} {\emph {Bose-{Einstein} {Condensation}
  of {Long}-{Lifetime} {Polaritons} in {Thermal} {Equilibrium}}},\ }\href
  {\doibase 10.1103/PhysRevLett.118.016602} {\bibfield  {journal} {\bibinfo
  {journal} {Physical Review Letters}\ }\textbf {\bibinfo {volume} {118}},\
  \bibinfo {pages} {016602} (\bibinfo {year} {2017})}\BibitemShut {NoStop}%
\bibitem [{\citenamefont {Roumpos}\ \emph {et~al.}(2012)\citenamefont
  {Roumpos}, \citenamefont {Lohse}, \citenamefont {Nitsche}, \citenamefont
  {Keeling}, \citenamefont {Szymanska}, \citenamefont {Littlewood},
  \citenamefont {Löffler}, \citenamefont {Hofling}, \citenamefont {Worschech},
  \citenamefont {Forchel},\ and\ \citenamefont
  {Yamamoto}}]{roumpos_power-law_2012}%
  \BibitemOpen
  \bibfield  {author} {\bibinfo {author} {\bibfnamefont {G.}~\bibnamefont
  {Roumpos}}, \bibinfo {author} {\bibfnamefont {M.}~\bibnamefont {Lohse}},
  \bibinfo {author} {\bibfnamefont {W.~H.}\ \bibnamefont {Nitsche}}, \bibinfo
  {author} {\bibfnamefont {J.}~\bibnamefont {Keeling}}, \bibinfo {author}
  {\bibfnamefont {M.~H.}\ \bibnamefont {Szymanska}}, \bibinfo {author}
  {\bibfnamefont {P.~B.}\ \bibnamefont {Littlewood}}, \bibinfo {author}
  {\bibfnamefont {A.}~\bibnamefont {Löffler}}, \bibinfo {author}
  {\bibfnamefont {S.}~\bibnamefont {Hofling}}, \bibinfo {author} {\bibfnamefont
  {L.}~\bibnamefont {Worschech}}, \bibinfo {author} {\bibfnamefont
  {A.}~\bibnamefont {Forchel}}, \ and\ \bibinfo {author} {\bibfnamefont
  {Y.}~\bibnamefont {Yamamoto}},\ }\bibfield  {title} {\bibinfo {title} {\emph
  {Power-law decay of the spatial correlation function in exciton-polariton
  condensates}},\ }\href {\doibase 10.1073/pnas.1107970109} {\bibfield
  {journal} {\bibinfo  {journal} {Proc. Natl. Acad. Sci.}\ }\textbf {\bibinfo
  {volume} {109}},\ \bibinfo {pages} {6467} (\bibinfo {year}
  {2012})}\BibitemShut {NoStop}%
\bibitem [{\citenamefont {Nitsche}\ \emph {et~al.}(2014)\citenamefont
  {Nitsche}, \citenamefont {Kim}, \citenamefont {Roumpos}, \citenamefont
  {Schneider}, \citenamefont {Kamp}, \citenamefont {Hofling}, \citenamefont
  {Forchel},\ and\ \citenamefont {Yamamoto}}]{nitsche_algebraic_2014}%
  \BibitemOpen
  \bibfield  {author} {\bibinfo {author} {\bibfnamefont {W.~H.}\ \bibnamefont
  {Nitsche}}, \bibinfo {author} {\bibfnamefont {N.~Y.}\ \bibnamefont {Kim}},
  \bibinfo {author} {\bibfnamefont {G.}~\bibnamefont {Roumpos}}, \bibinfo
  {author} {\bibfnamefont {C.}~\bibnamefont {Schneider}}, \bibinfo {author}
  {\bibfnamefont {M.}~\bibnamefont {Kamp}}, \bibinfo {author} {\bibfnamefont
  {S.}~\bibnamefont {Hofling}}, \bibinfo {author} {\bibfnamefont
  {A.}~\bibnamefont {Forchel}}, \ and\ \bibinfo {author} {\bibfnamefont
  {Y.}~\bibnamefont {Yamamoto}},\ }\bibfield  {title} {\bibinfo {title} {\emph
  {Algebraic order and the {Berezinskii}-{Kosterlitz}-{Thouless} transition in
  an exciton-polariton gas}},\ }\href {\doibase 10.1103/PhysRevB.90.205430}
  {\bibfield  {journal} {\bibinfo  {journal} {Physical Review B}\ }\textbf
  {\bibinfo {volume} {90}},\ \bibinfo {pages} {205430} (\bibinfo {year}
  {2014})}\BibitemShut {NoStop}%
\bibitem [{\citenamefont {Baboux}\ \emph {et~al.}(2018)\citenamefont {Baboux},
  \citenamefont {Bernardis}, \citenamefont {Goblot}, \citenamefont {Gladilin},
  \citenamefont {Gomez}, \citenamefont {Galopin}, \citenamefont {Gratiet},
  \citenamefont {Lemaître}, \citenamefont {Sagnes}, \citenamefont {Carusotto},
  \citenamefont {Wouters}, \citenamefont {Amo},\ and\ \citenamefont
  {Bloch}}]{baboux_unstable_2018}%
  \BibitemOpen
  \bibfield  {author} {\bibinfo {author} {\bibfnamefont {F.}~\bibnamefont
  {Baboux}}, \bibinfo {author} {\bibfnamefont {D.~D.}\ \bibnamefont
  {Bernardis}}, \bibinfo {author} {\bibfnamefont {V.}~\bibnamefont {Goblot}},
  \bibinfo {author} {\bibfnamefont {V.~N.}\ \bibnamefont {Gladilin}}, \bibinfo
  {author} {\bibfnamefont {C.}~\bibnamefont {Gomez}}, \bibinfo {author}
  {\bibfnamefont {E.}~\bibnamefont {Galopin}}, \bibinfo {author} {\bibfnamefont
  {L.~L.}\ \bibnamefont {Gratiet}}, \bibinfo {author} {\bibfnamefont
  {A.}~\bibnamefont {Lemaître}}, \bibinfo {author} {\bibfnamefont
  {I.}~\bibnamefont {Sagnes}}, \bibinfo {author} {\bibfnamefont
  {I.}~\bibnamefont {Carusotto}}, \bibinfo {author} {\bibfnamefont
  {M.}~\bibnamefont {Wouters}}, \bibinfo {author} {\bibfnamefont
  {A.}~\bibnamefont {Amo}}, \ and\ \bibinfo {author} {\bibfnamefont
  {J.}~\bibnamefont {Bloch}},\ }\bibfield  {title} {\bibinfo {title} {\emph
  {Unstable and stable regimes of polariton condensation}},\ }\href {\doibase
  10.1364/OPTICA.5.001163} {\bibfield  {journal} {\bibinfo  {journal} {Optica}\
  }\textbf {\bibinfo {volume} {5}},\ \bibinfo {pages} {1163} (\bibinfo {year}
  {2018})}\BibitemShut {NoStop}%
\bibitem [{\citenamefont {Love}\ \emph {et~al.}(2008)\citenamefont {Love},
  \citenamefont {Krizhanovskii}, \citenamefont {Whittaker}, \citenamefont
  {Bouchekioua}, \citenamefont {Sanvitto}, \citenamefont {Rizeiqi},
  \citenamefont {Bradley}, \citenamefont {Skolnick}, \citenamefont {Eastham},
  \citenamefont {Andre},\ and\ \citenamefont {Dang}}]{love_intrinsic_2008}%
  \BibitemOpen
  \bibfield  {author} {\bibinfo {author} {\bibfnamefont {A.~P.~D.}\
  \bibnamefont {Love}}, \bibinfo {author} {\bibfnamefont {D.~N.}\ \bibnamefont
  {Krizhanovskii}}, \bibinfo {author} {\bibfnamefont {D.~M.}\ \bibnamefont
  {Whittaker}}, \bibinfo {author} {\bibfnamefont {R.}~\bibnamefont
  {Bouchekioua}}, \bibinfo {author} {\bibfnamefont {D.}~\bibnamefont
  {Sanvitto}}, \bibinfo {author} {\bibfnamefont {S.~A.}\ \bibnamefont
  {Rizeiqi}}, \bibinfo {author} {\bibfnamefont {R.}~\bibnamefont {Bradley}},
  \bibinfo {author} {\bibfnamefont {M.~S.}\ \bibnamefont {Skolnick}}, \bibinfo
  {author} {\bibfnamefont {P.~R.}\ \bibnamefont {Eastham}}, \bibinfo {author}
  {\bibfnamefont {R.}~\bibnamefont {Andre}}, \ and\ \bibinfo {author}
  {\bibfnamefont {L.~S.}\ \bibnamefont {Dang}},\ }\bibfield  {title} {\bibinfo
  {title} {\emph {Intrinsic {Decoherence} {Mechanisms} in the {Microcavity}
  {Polariton} {Condensate}}},\ }\href {\doibase 10.1103/PhysRevLett.101.067404}
  {\bibfield  {journal} {\bibinfo  {journal} {Physical Review Letters}\
  }\textbf {\bibinfo {volume} {101}},\ \bibinfo {pages} {067404} (\bibinfo
  {year} {2008})}\BibitemShut {NoStop}%
\bibitem [{\citenamefont {Haug}\ \emph {et~al.}(2012)\citenamefont {Haug},
  \citenamefont {Doan}, \citenamefont {Cao},\ and\ \citenamefont
  {Thoai}}]{haug_temporal_2012}%
  \BibitemOpen
  \bibfield  {author} {\bibinfo {author} {\bibfnamefont {H.}~\bibnamefont
  {Haug}}, \bibinfo {author} {\bibfnamefont {T.~D.}\ \bibnamefont {Doan}},
  \bibinfo {author} {\bibfnamefont {H.~T.}\ \bibnamefont {Cao}}, \ and\
  \bibinfo {author} {\bibfnamefont {D.~B.~T.}\ \bibnamefont {Thoai}},\
  }\bibfield  {title} {\bibinfo {title} {\emph {Temporal first- and
  second-order correlations in a polariton condensate}},\ }\href {\doibase
  10.1103/PhysRevB.85.205310} {\bibfield  {journal} {\bibinfo  {journal}
  {Physical Review B}\ }\textbf {\bibinfo {volume} {85}},\ \bibinfo {pages}
  {205310} (\bibinfo {year} {2012})}\BibitemShut {NoStop}%
\bibitem [{\citenamefont {Whittaker}\ and\ \citenamefont
  {Eastham}(2009)}]{whittaker_coherence_2009}%
  \BibitemOpen
  \bibfield  {author} {\bibinfo {author} {\bibfnamefont {D.~M.}\ \bibnamefont
  {Whittaker}}\ and\ \bibinfo {author} {\bibfnamefont {P.~R.}\ \bibnamefont
  {Eastham}},\ }\bibfield  {title} {\bibinfo {title} {\emph {Coherence
  properties of the microcavity polariton condensate}},\ }\href {\doibase
  10.1209/0295-5075/87/27002} {\bibfield  {journal} {\bibinfo  {journal} {EPL
  (Europhysics Letters)}\ }\textbf {\bibinfo {volume} {87}},\ \bibinfo {pages}
  {27002} (\bibinfo {year} {2009})}\BibitemShut {NoStop}%
\bibitem [{\citenamefont {Spano}\ \emph {et~al.}(2013)\citenamefont {Spano},
  \citenamefont {Cuadra}, \citenamefont {Lingg}, \citenamefont {Sanvitto},
  \citenamefont {Martin}, \citenamefont {Eastham}, \citenamefont {Poel},
  \citenamefont {Hvam},\ and\ \citenamefont {Viña}}]{spano_build_2013}%
  \BibitemOpen
  \bibfield  {author} {\bibinfo {author} {\bibfnamefont {R.}~\bibnamefont
  {Spano}}, \bibinfo {author} {\bibfnamefont {J.}~\bibnamefont {Cuadra}},
  \bibinfo {author} {\bibfnamefont {C.}~\bibnamefont {Lingg}}, \bibinfo
  {author} {\bibfnamefont {D.}~\bibnamefont {Sanvitto}}, \bibinfo {author}
  {\bibfnamefont {M.~D.}\ \bibnamefont {Martin}}, \bibinfo {author}
  {\bibfnamefont {P.~R.}\ \bibnamefont {Eastham}}, \bibinfo {author}
  {\bibfnamefont {M.~v.~d.}\ \bibnamefont {Poel}}, \bibinfo {author}
  {\bibfnamefont {J.~M.}\ \bibnamefont {Hvam}}, \ and\ \bibinfo {author}
  {\bibfnamefont {L.}~\bibnamefont {Viña}},\ }\bibfield  {title} {\bibinfo
  {title} {\emph {Build up of off-diagonal long-range order in microcavity
  exciton-polaritons across the parametric threshold}},\ }\href {\doibase
  10.1364/OE.21.010792} {\bibfield  {journal} {\bibinfo  {journal} {Optics
  Express}\ }\textbf {\bibinfo {volume} {21}},\ \bibinfo {pages} {10792}
  (\bibinfo {year} {2013})}\BibitemShut {NoStop}%
\bibitem [{\citenamefont {Clauset}\ \emph {et~al.}(2009)\citenamefont
  {Clauset}, \citenamefont {Shalizi},\ and\ \citenamefont
  {Newman}}]{clauset_power-law_2009}%
  \BibitemOpen
  \bibfield  {author} {\bibinfo {author} {\bibfnamefont {A.}~\bibnamefont
  {Clauset}}, \bibinfo {author} {\bibfnamefont {C.~R.}\ \bibnamefont
  {Shalizi}}, \ and\ \bibinfo {author} {\bibfnamefont {M.~E.~J.}\ \bibnamefont
  {Newman}},\ }\bibfield  {title} {\bibinfo {title} {\emph {Power-{Law}
  {Distributions} in {Empirical} {Data}}},\ }\href {\doibase 10.1137/070710111}
  {\bibfield  {journal} {\bibinfo  {journal} {SIAM Review}\ }\textbf {\bibinfo
  {volume} {51}},\ \bibinfo {pages} {661} (\bibinfo {year} {2009})}\BibitemShut
  {NoStop}%
\bibitem [{\citenamefont {De~Giorgi}\ \emph {et~al.}(2018)\citenamefont
  {De~Giorgi}, \citenamefont {Ramezani}, \citenamefont {Todisco}, \citenamefont
  {Halpin}, \citenamefont {Caputo}, \citenamefont {Fieramosca}, \citenamefont
  {{Gomez-Rivas}},\ and\ \citenamefont
  {Sanvitto}}]{de_giorgi_interaction_2018}%
  \BibitemOpen
  \bibfield  {author} {\bibinfo {author} {\bibfnamefont {M.}~\bibnamefont
  {De~Giorgi}}, \bibinfo {author} {\bibfnamefont {M.}~\bibnamefont {Ramezani}},
  \bibinfo {author} {\bibfnamefont {F.}~\bibnamefont {Todisco}}, \bibinfo
  {author} {\bibfnamefont {A.}~\bibnamefont {Halpin}}, \bibinfo {author}
  {\bibfnamefont {D.}~\bibnamefont {Caputo}}, \bibinfo {author} {\bibfnamefont
  {A.}~\bibnamefont {Fieramosca}}, \bibinfo {author} {\bibfnamefont
  {J.}~\bibnamefont {{Gomez-Rivas}}}, \ and\ \bibinfo {author} {\bibfnamefont
  {D.}~\bibnamefont {Sanvitto}},\ }\bibfield  {title} {\bibinfo {title} {\emph
  {Interaction and {{Coherence}} of a {{Plasmon}}\textendash{{Exciton Polariton
  Condensate}}}},\ }\href {\doibase 10.1021/acsphotonics.8b00662} {\bibfield
  {journal} {\bibinfo  {journal} {ACS Photonics}\ }\textbf {\bibinfo {volume}
  {5}},\ \bibinfo {pages} {3666} (\bibinfo {year} {2018})}\BibitemShut
  {NoStop}%
\bibitem [{Note1()}]{Note1}%
  \BibitemOpen
  \bibinfo {note} {The condensate size \SI {500}{\micro m} is 20 times the
  spatial coherence length of the polaritons in the uncondensed system (below
  the first threshold, \SI {22}{\micro m}~\cite {SM}) and 100 times the healing
  length $\xi _0 =\protect \hbar /\protect \sqrt {{2m_\protect \mathrm
  {eff}gn}}=0.6$...\SI {6}{\micro m}, where $m_\protect \mathrm
  {eff}=1e^{-7}-1e^{-5} m_\protect \mathrm {e}$ and $gn=0.02$~eV~\cite
  {Vakevainen2020}.}\BibitemShut {Stop}%
\bibitem [{\citenamefont {Hakala}\ \emph {et~al.}(2018)\citenamefont {Hakala},
  \citenamefont {Moilanen}, \citenamefont {V{\"a}kev{\"a}inen}, \citenamefont
  {Guo}, \citenamefont {Martikainen}, \citenamefont {Daskalakis}, \citenamefont
  {Rekola}, \citenamefont {Julku},\ and\ \citenamefont
  {T{\"o}rm{\"a}}}]{hakala_bose-einstein_2018}%
  \BibitemOpen
  \bibfield  {author} {\bibinfo {author} {\bibfnamefont {T.~K.}\ \bibnamefont
  {Hakala}}, \bibinfo {author} {\bibfnamefont {A.~J.}\ \bibnamefont
  {Moilanen}}, \bibinfo {author} {\bibfnamefont {A.~I.}\ \bibnamefont
  {V{\"a}kev{\"a}inen}}, \bibinfo {author} {\bibfnamefont {R.}~\bibnamefont
  {Guo}}, \bibinfo {author} {\bibfnamefont {J.-P.}\ \bibnamefont
  {Martikainen}}, \bibinfo {author} {\bibfnamefont {K.~S.}\ \bibnamefont
  {Daskalakis}}, \bibinfo {author} {\bibfnamefont {H.~T.}\ \bibnamefont
  {Rekola}}, \bibinfo {author} {\bibfnamefont {A.}~\bibnamefont {Julku}}, \
  and\ \bibinfo {author} {\bibfnamefont {P.}~\bibnamefont {T{\"o}rm{\"a}}},\
  }\bibfield  {title} {\bibinfo {title} {\emph {Bose-{{Einstein}} Condensation
  in a Plasmonic Lattice}},\ }\href {\doibase 10.1038/s41567-018-0109-9}
  {\bibfield  {journal} {\bibinfo  {journal} {Nature Physics}\ }\textbf
  {\bibinfo {volume} {14}},\ \bibinfo {pages} {739} (\bibinfo {year}
  {2018})}\BibitemShut {NoStop}%
\bibitem [{\citenamefont {V{\"a}kev{\"a}inen}\ \emph
  {et~al.}(2020)\citenamefont {V{\"a}kev{\"a}inen}, \citenamefont {Moilanen},
  \citenamefont {Ne{\v{c}}ada}, \citenamefont {Hakala}, \citenamefont
  {Daskalakis},\ and\ \citenamefont {T{\"o}rm{\"a}}}]{Vakevainen2020}%
  \BibitemOpen
  \bibfield  {author} {\bibinfo {author} {\bibfnamefont {A.~I.}\ \bibnamefont
  {V{\"a}kev{\"a}inen}}, \bibinfo {author} {\bibfnamefont {A.~J.}\ \bibnamefont
  {Moilanen}}, \bibinfo {author} {\bibfnamefont {M.}~\bibnamefont
  {Ne{\v{c}}ada}}, \bibinfo {author} {\bibfnamefont {T.~K.}\ \bibnamefont
  {Hakala}}, \bibinfo {author} {\bibfnamefont {K.~S.}\ \bibnamefont
  {Daskalakis}}, \ and\ \bibinfo {author} {\bibfnamefont {P.}~\bibnamefont
  {T{\"o}rm{\"a}}},\ }\bibfield  {title} {\bibinfo {title} {\emph
  {Sub-picosecond thermalization dynamics in condensation of strongly coupled
  lattice plasmons}},\ }\href {\doibase 10.1038/s41467-020-16906-1} {\bibfield
  {journal} {\bibinfo  {journal} {Nature Communications}\ }\textbf {\bibinfo
  {volume} {11}},\ \bibinfo {pages} {3139} (\bibinfo {year}
  {2020})}\BibitemShut {NoStop}%
\bibitem [{\citenamefont {Wang}\ \emph
  {et~al.}(2018{\natexlab{a}})\citenamefont {Wang}, \citenamefont {Ramezani},
  \citenamefont {V{\"a}kev{\"a}inen}, \citenamefont {T{\"o}rm{\"a}},
  \citenamefont {Rivas},\ and\ \citenamefont {Odom}}]{wang_rich_2018}%
  \BibitemOpen
  \bibfield  {author} {\bibinfo {author} {\bibfnamefont {W.}~\bibnamefont
  {Wang}}, \bibinfo {author} {\bibfnamefont {M.}~\bibnamefont {Ramezani}},
  \bibinfo {author} {\bibfnamefont {A.~I.}\ \bibnamefont {V{\"a}kev{\"a}inen}},
  \bibinfo {author} {\bibfnamefont {P.}~\bibnamefont {T{\"o}rm{\"a}}}, \bibinfo
  {author} {\bibfnamefont {J.~G.}\ \bibnamefont {Rivas}}, \ and\ \bibinfo
  {author} {\bibfnamefont {T.~W.}\ \bibnamefont {Odom}},\ }\bibfield  {title}
  {\bibinfo {title} {\emph {The Rich Photonic World of Plasmonic Nanoparticle
  Arrays}},\ }\href {\doibase 10.1016/j.mattod.2017.09.002} {\bibfield
  {journal} {\bibinfo  {journal} {Materials Today}\ }\textbf {\bibinfo {volume}
  {21}},\ \bibinfo {pages} {303} (\bibinfo {year}
  {2018}{\natexlab{a}})}\BibitemShut {NoStop}%
\bibitem [{\citenamefont {Kravets}\ \emph {et~al.}(2018)\citenamefont
  {Kravets}, \citenamefont {Kabashin}, \citenamefont {Barnes},\ and\
  \citenamefont {Grigorenko}}]{kravets_plasmonic_2018}%
  \BibitemOpen
  \bibfield  {author} {\bibinfo {author} {\bibfnamefont {V.~G.}\ \bibnamefont
  {Kravets}}, \bibinfo {author} {\bibfnamefont {A.~V.}\ \bibnamefont
  {Kabashin}}, \bibinfo {author} {\bibfnamefont {W.~L.}\ \bibnamefont
  {Barnes}}, \ and\ \bibinfo {author} {\bibfnamefont {A.~N.}\ \bibnamefont
  {Grigorenko}},\ }\bibfield  {title} {\bibinfo {title} {\emph {Plasmonic
  {{Surface Lattice Resonances}}: {{A Review}} of {{Properties}} and
  {{Applications}}}},\ }\href {\doibase 10.1021/acs.chemrev.8b00243} {\bibfield
   {journal} {\bibinfo  {journal} {Chemical Reviews}\ }\textbf {\bibinfo
  {volume} {118}},\ \bibinfo {pages} {5912} (\bibinfo {year}
  {2018})}\BibitemShut {NoStop}%
\bibitem [{\citenamefont {Wang}\ \emph
  {et~al.}(2018{\natexlab{b}})\citenamefont {Wang}, \citenamefont {Wang},
  \citenamefont {Knudson}, \citenamefont {Schatz},\ and\ \citenamefont
  {Odom}}]{wang_structural_2018}%
  \BibitemOpen
  \bibfield  {author} {\bibinfo {author} {\bibfnamefont {D.}~\bibnamefont
  {Wang}}, \bibinfo {author} {\bibfnamefont {W.}~\bibnamefont {Wang}}, \bibinfo
  {author} {\bibfnamefont {M.~P.}\ \bibnamefont {Knudson}}, \bibinfo {author}
  {\bibfnamefont {G.~C.}\ \bibnamefont {Schatz}}, \ and\ \bibinfo {author}
  {\bibfnamefont {T.~W.}\ \bibnamefont {Odom}},\ }\bibfield  {title} {\bibinfo
  {title} {\emph {Structural {{Engineering}} in {{Plasmon Nanolasers}}}},\
  }\href {\doibase 10.1021/acs.chemrev.7b00424} {\bibfield  {journal} {\bibinfo
   {journal} {Chemical Reviews}\ }\textbf {\bibinfo {volume} {118}},\ \bibinfo
  {pages} {2865} (\bibinfo {year} {2018}{\natexlab{b}})}\BibitemShut {NoStop}%
\bibitem [{\citenamefont {Hakala}\ \emph {et~al.}(2017)\citenamefont {Hakala},
  \citenamefont {Rekola}, \citenamefont {V{\"a}kev{\"a}inen}, \citenamefont
  {Martikainen}, \citenamefont {Ne{\v c}ada}, \citenamefont {Moilanen},\ and\
  \citenamefont {T{\"o}rm{\"a}}}]{hakala_lasing_2017}%
  \BibitemOpen
  \bibfield  {author} {\bibinfo {author} {\bibfnamefont {T.~K.}\ \bibnamefont
  {Hakala}}, \bibinfo {author} {\bibfnamefont {H.~T.}\ \bibnamefont {Rekola}},
  \bibinfo {author} {\bibfnamefont {A.~I.}\ \bibnamefont {V{\"a}kev{\"a}inen}},
  \bibinfo {author} {\bibfnamefont {J.-P.}\ \bibnamefont {Martikainen}},
  \bibinfo {author} {\bibfnamefont {M.}~\bibnamefont {Ne{\v c}ada}}, \bibinfo
  {author} {\bibfnamefont {A.~J.}\ \bibnamefont {Moilanen}}, \ and\ \bibinfo
  {author} {\bibfnamefont {P.}~\bibnamefont {T{\"o}rm{\"a}}},\ }\bibfield
  {title} {\bibinfo {title} {\emph {Lasing in Dark and Bright Modes of a
  Finite-Sized Plasmonic Lattice}},\ }\href {\doibase 10.1038/ncomms13687}
  {\bibfield  {journal} {\bibinfo  {journal} {Nature Communications}\ }\textbf
  {\bibinfo {volume} {8}},\ \bibinfo {pages} {13687} (\bibinfo {year}
  {2017})}\BibitemShut {NoStop}%
\bibitem [{\citenamefont {Daskalakis}\ \emph {et~al.}(2018)\citenamefont
  {Daskalakis}, \citenamefont {V{\"a}kev{\"a}inen}, \citenamefont
  {Martikainen}, \citenamefont {Hakala},\ and\ \citenamefont
  {T{\"o}rm{\"a}}}]{daskalakis_ultrafast_2018}%
  \BibitemOpen
  \bibfield  {author} {\bibinfo {author} {\bibfnamefont {K.~S.}\ \bibnamefont
  {Daskalakis}}, \bibinfo {author} {\bibfnamefont {A.~I.}\ \bibnamefont
  {V{\"a}kev{\"a}inen}}, \bibinfo {author} {\bibfnamefont {J.-P.}\ \bibnamefont
  {Martikainen}}, \bibinfo {author} {\bibfnamefont {T.~K.}\ \bibnamefont
  {Hakala}}, \ and\ \bibinfo {author} {\bibfnamefont {P.}~\bibnamefont
  {T{\"o}rm{\"a}}},\ }\bibfield  {title} {\bibinfo {title} {\emph {Ultrafast
  {{Pulse Generation}} in an {{Organic Nanoparticle}}-{{Array Laser}}}},\
  }\href {\doibase 10.1021/acs.nanolett.8b00531} {\bibfield  {journal}
  {\bibinfo  {journal} {Nano Letters}\ }\textbf {\bibinfo {volume} {18}},\
  \bibinfo {pages} {2658} (\bibinfo {year} {2018})}\BibitemShut {NoStop}%
\bibitem [{\citenamefont {Moilanen}\ \emph {et~al.}(2021)\citenamefont
  {Moilanen}, \citenamefont {Konstantinos}, \citenamefont {Taskinen},\ and\
  \citenamefont {T\"orm\"a}}]{SM}%
  \BibitemOpen
  \bibfield  {author} {\bibinfo {author} {\bibfnamefont {A.~J.}\ \bibnamefont
  {Moilanen}}, \bibinfo {author} {\bibfnamefont {S.~D.}\ \bibnamefont
  {Konstantinos}}, \bibinfo {author} {\bibfnamefont {J.~M.}\ \bibnamefont
  {Taskinen}}, \ and\ \bibinfo {author} {\bibfnamefont {P.}~\bibnamefont
  {T\"orm\"a}},\ }\href@noop {} {\bibinfo {title} {\emph {Supplemental
  Material}}} (\bibinfo {year} {2021}),\ \bibinfo {note} {see Supplemental
  Material for detailed methodology and additional experiments, which includes
  Refs.~\cite{johnson_optical_1972,torma_strong_2015,daskalakis_room-temperature_2014,Born1998}.}\BibitemShut
  {Stop}%
\bibitem [{Note2()}]{Note2}%
  \BibitemOpen
  \bibinfo {note} {In this case, coherences may be present in the system
  density matrix even if the photon populations followed a thermal
  distribution.}\BibitemShut {Stop}%
\bibitem [{\citenamefont {Hoang}\ \emph {et~al.}(2017)\citenamefont {Hoang},
  \citenamefont {Akselrod}, \citenamefont {Yang}, \citenamefont {Odom},\ and\
  \citenamefont {Mikkelsen}}]{hoang_millimeter-scale_2017}%
  \BibitemOpen
  \bibfield  {author} {\bibinfo {author} {\bibfnamefont {T.~B.}\ \bibnamefont
  {Hoang}}, \bibinfo {author} {\bibfnamefont {G.~M.}\ \bibnamefont {Akselrod}},
  \bibinfo {author} {\bibfnamefont {A.}~\bibnamefont {Yang}}, \bibinfo {author}
  {\bibfnamefont {T.~W.}\ \bibnamefont {Odom}}, \ and\ \bibinfo {author}
  {\bibfnamefont {M.~H.}\ \bibnamefont {Mikkelsen}},\ }\bibfield  {title}
  {\bibinfo {title} {\emph {Millimeter-{Scale} {Spatial} {Coherence} from a
  {Plasmon} {Laser}}},\ }\href {\doibase 10.1021/acs.nanolett.7b02677}
  {\bibfield  {journal} {\bibinfo  {journal} {Nano Letters}\ }\textbf {\bibinfo
  {volume} {17}},\ \bibinfo {pages} {6690} (\bibinfo {year}
  {2017})}\BibitemShut {NoStop}%
\bibitem [{\citenamefont {Kim}\ \emph {et~al.}(2016)\citenamefont {Kim},
  \citenamefont {Zhang}, \citenamefont {Wang}, \citenamefont {Fischer},
  \citenamefont {Brodbeck}, \citenamefont {Kamp}, \citenamefont {Schneider},
  \citenamefont {Hofling},\ and\ \citenamefont {Deng}}]{kim_coherent_2016}%
  \BibitemOpen
  \bibfield  {author} {\bibinfo {author} {\bibfnamefont {S.}~\bibnamefont
  {Kim}}, \bibinfo {author} {\bibfnamefont {B.}~\bibnamefont {Zhang}}, \bibinfo
  {author} {\bibfnamefont {Z.}~\bibnamefont {Wang}}, \bibinfo {author}
  {\bibfnamefont {J.}~\bibnamefont {Fischer}}, \bibinfo {author} {\bibfnamefont
  {S.}~\bibnamefont {Brodbeck}}, \bibinfo {author} {\bibfnamefont
  {M.}~\bibnamefont {Kamp}}, \bibinfo {author} {\bibfnamefont {C.}~\bibnamefont
  {Schneider}}, \bibinfo {author} {\bibfnamefont {S.}~\bibnamefont {Hofling}},
  \ and\ \bibinfo {author} {\bibfnamefont {H.}~\bibnamefont {Deng}},\
  }\bibfield  {title} {\bibinfo {title} {\emph {Coherent {Polariton}
  {Laser}}},\ }\href {\doibase 10.1103/PhysRevX.6.011026} {\bibfield  {journal}
  {\bibinfo  {journal} {Physical Review X}\ }\textbf {\bibinfo {volume} {6}},\
  \bibinfo {pages} {011026} (\bibinfo {year} {2016})}\BibitemShut {NoStop}%
\bibitem [{\citenamefont {Halpin-Healy}\ and\ \citenamefont
  {Palasantzas}(2014)}]{halpin-healy_universal_2014}%
  \BibitemOpen
  \bibfield  {author} {\bibinfo {author} {\bibfnamefont {T.}~\bibnamefont
  {Halpin-Healy}}\ and\ \bibinfo {author} {\bibfnamefont {G.}~\bibnamefont
  {Palasantzas}},\ }\bibfield  {title} {\bibinfo {title} {\emph {Universal
  correlators and distributions as experimental signatures of (2+1)-dimensional
  {Kardar}-{Parisi}-{Zhang} growth}},\ }\href {\doibase
  10.1209/0295-5075/105/50001} {\bibfield  {journal} {\bibinfo  {journal} {EPL
  (Europhysics Letters)}\ }\textbf {\bibinfo {volume} {105}},\ \bibinfo {pages}
  {50001} (\bibinfo {year} {2014})}\BibitemShut {NoStop}%
\bibitem [{\citenamefont {Pagnani}\ and\ \citenamefont
  {Parisi}(2015)}]{pagnani_numerical_2015}%
  \BibitemOpen
  \bibfield  {author} {\bibinfo {author} {\bibfnamefont {A.}~\bibnamefont
  {Pagnani}}\ and\ \bibinfo {author} {\bibfnamefont {G.}~\bibnamefont
  {Parisi}},\ }\bibfield  {title} {\bibinfo {title} {\emph {Numerical estimate
  of the {Kardar}-{Parisi}-{Zhang} universality class in (2+1) dimensions}},\
  }\href {\doibase 10.1103/PhysRevE.92.010101} {\bibfield  {journal} {\bibinfo
  {journal} {Physical Review E}\ }\textbf {\bibinfo {volume} {92}},\ \bibinfo
  {pages} {010101} (\bibinfo {year} {2015})}\BibitemShut {NoStop}%
\bibitem [{\citenamefont {Miranda}\ and\ \citenamefont
  {Aarao~Reis}(2008)}]{miranda_numerical_2008}%
  \BibitemOpen
  \bibfield  {author} {\bibinfo {author} {\bibfnamefont {V.~G.}\ \bibnamefont
  {Miranda}}\ and\ \bibinfo {author} {\bibfnamefont {F.~D.~A.}\ \bibnamefont
  {Aarao~Reis}},\ }\bibfield  {title} {\bibinfo {title} {\emph {Numerical study
  of the {Kardar}-{Parisi}-{Zhang} equation}},\ }\href {\doibase
  10.1103/PhysRevE.77.031134} {\bibfield  {journal} {\bibinfo  {journal}
  {Physical Review E}\ }\textbf {\bibinfo {volume} {77}},\ \bibinfo {pages}
  {031134} (\bibinfo {year} {2008})}\BibitemShut {NoStop}%
\bibitem [{\citenamefont {He}\ \emph {et~al.}(2015)\citenamefont {He},
  \citenamefont {Sieberer}, \citenamefont {Altman},\ and\ \citenamefont
  {Diehl}}]{he_scaling_2015}%
  \BibitemOpen
  \bibfield  {author} {\bibinfo {author} {\bibfnamefont {L.}~\bibnamefont
  {He}}, \bibinfo {author} {\bibfnamefont {L.~M.}\ \bibnamefont {Sieberer}},
  \bibinfo {author} {\bibfnamefont {E.}~\bibnamefont {Altman}}, \ and\ \bibinfo
  {author} {\bibfnamefont {S.}~\bibnamefont {Diehl}},\ }\bibfield  {title}
  {\bibinfo {title} {\emph {Scaling properties of one-dimensional
  driven-dissipative condensates}},\ }\href {\doibase
  10.1103/PhysRevB.92.155307} {\bibfield  {journal} {\bibinfo  {journal}
  {Physical Review B}\ }\textbf {\bibinfo {volume} {92}},\ \bibinfo {pages}
  {155307} (\bibinfo {year} {2015})}\BibitemShut {NoStop}%
\bibitem [{\citenamefont {Moerland}\ \emph {et~al.}(2017)\citenamefont
  {Moerland}, \citenamefont {Hakala}, \citenamefont {Martikainen},
  \citenamefont {Rekola}, \citenamefont {V{\"a}kev{\"a}inen},\ and\
  \citenamefont {T{\"o}rm{\"a}}}]{Moerland2017}%
  \BibitemOpen
  \bibfield  {author} {\bibinfo {author} {\bibfnamefont {R.}~\bibnamefont
  {Moerland}}, \bibinfo {author} {\bibfnamefont {T.}~\bibnamefont {Hakala}},
  \bibinfo {author} {\bibfnamefont {J.-P.}\ \bibnamefont {Martikainen}},
  \bibinfo {author} {\bibfnamefont {H.}~\bibnamefont {Rekola}}, \bibinfo
  {author} {\bibfnamefont {A.}~\bibnamefont {V{\"a}kev{\"a}inen}}, \ and\
  \bibinfo {author} {\bibfnamefont {P.}~\bibnamefont {T{\"o}rm{\"a}}},\
  }\bibfield  {title} {\bibinfo {title} {\emph {{Strong Coupling Between
  Organic Molecules and Plasmonic Nanostructures}}},\ }in\ \href@noop {} {\emph
  {\bibinfo {booktitle} {{Quantum Plasmonics}}}}\ (\bibinfo  {publisher}
  {{Springer, Cham}},\ \bibinfo {year} {2017})\BibitemShut {NoStop}%
\bibitem [{\citenamefont {Guo}\ \emph {et~al.}(2017)\citenamefont {Guo},
  \citenamefont {Hakala},\ and\ \citenamefont {Törmä}}]{guo_geometry_2017}%
  \BibitemOpen
  \bibfield  {author} {\bibinfo {author} {\bibfnamefont {R.}~\bibnamefont
  {Guo}}, \bibinfo {author} {\bibfnamefont {T.~K.}\ \bibnamefont {Hakala}}, \
  and\ \bibinfo {author} {\bibfnamefont {P.}~\bibnamefont {Törmä}},\
  }\bibfield  {title} {\bibinfo {title} {\emph {Geometry dependence of surface
  lattice resonances in plasmonic nanoparticle arrays}},\ }\href {\doibase
  10.1103/PhysRevB.95.155423} {\bibfield  {journal} {\bibinfo  {journal}
  {Physical Review B}\ }\textbf {\bibinfo {volume} {95}},\ \bibinfo {pages}
  {155423} (\bibinfo {year} {2017})}\BibitemShut {NoStop}%
\bibitem [{\citenamefont {Keeling}\ \emph {et~al.}(2017)\citenamefont
  {Keeling}, \citenamefont {Sieberer}, \citenamefont {Altman}, \citenamefont
  {Chen}, \citenamefont {Diehl},\ and\ \citenamefont
  {Toner}}]{keeling_superfluidity_2017}%
  \BibitemOpen
  \bibfield  {author} {\bibinfo {author} {\bibfnamefont {J.}~\bibnamefont
  {Keeling}}, \bibinfo {author} {\bibfnamefont {L.~M.}\ \bibnamefont
  {Sieberer}}, \bibinfo {author} {\bibfnamefont {E.}~\bibnamefont {Altman}},
  \bibinfo {author} {\bibfnamefont {L.}~\bibnamefont {Chen}}, \bibinfo {author}
  {\bibfnamefont {S.}~\bibnamefont {Diehl}}, \ and\ \bibinfo {author}
  {\bibfnamefont {J.}~\bibnamefont {Toner}},\ }\bibfield  {title} {\bibinfo
  {title} {\emph {Superfluidity and {{Phase Correlations}} of {{Driven
  Dissipative Condensates}}}},\ }in\ \href@noop {} {\emph {\bibinfo {booktitle}
  {Universal {{Themes}} of {{Bose}}-{{Einstein Condensation}}}}}\ (\bibinfo
  {publisher} {{Cambridge University Press}},\ \bibinfo {year} {2017})\ pp.\
  \bibinfo {pages} {205--230},\ \Eprint {http://arxiv.org/abs/1601.04495}
  {arXiv:1601.04495} \BibitemShut {NoStop}%
\bibitem [{\citenamefont {Johnson}\ and\ \citenamefont
  {Christy}(1972)}]{johnson_optical_1972}%
  \BibitemOpen
  \bibfield  {author} {\bibinfo {author} {\bibfnamefont {P.~B.}\ \bibnamefont
  {Johnson}}\ and\ \bibinfo {author} {\bibfnamefont {R.~W.}\ \bibnamefont
  {Christy}},\ }\bibfield  {title} {\bibinfo {title} {\emph {Optical constants
  of noble metals}},\ }\href@noop {} {\bibfield  {journal} {\bibinfo  {journal}
  {Physical Review E}\ }\textbf {\bibinfo {volume} {6}},\ \bibinfo {pages}
  {4370} (\bibinfo {year} {1972})}\BibitemShut {NoStop}%
\bibitem [{\citenamefont {T{\"o}rm{\"a}}\ and\ \citenamefont
  {Barnes}(2015)}]{torma_strong_2015}%
  \BibitemOpen
  \bibfield  {author} {\bibinfo {author} {\bibfnamefont {P.}~\bibnamefont
  {T{\"o}rm{\"a}}}\ and\ \bibinfo {author} {\bibfnamefont {W.~L.}\ \bibnamefont
  {Barnes}},\ }\bibfield  {title} {\bibinfo {title} {\emph {Strong Coupling
  between Surface Plasmon Polaritons and Emitters: A Review}},\ }\href
  {\doibase 10.1088/0034-4885/78/1/013901} {\bibfield  {journal} {\bibinfo
  {journal} {Reports on Progress in Physics}\ }\textbf {\bibinfo {volume}
  {78}},\ \bibinfo {pages} {013901} (\bibinfo {year} {2015})}\BibitemShut
  {NoStop}%
\bibitem [{\citenamefont
  {Daskalakis}(2014)}]{daskalakis_room-temperature_2014}%
  \BibitemOpen
  \bibfield  {author} {\bibinfo {author} {\bibfnamefont {K.~S.}\ \bibnamefont
  {Daskalakis}},\ }\emph {\bibinfo {title} {Room-temperature polariton
  condensates in all-dielectric microcavities}},\ \href@noop {} {Ph.D.
  thesis},\ \bibinfo  {school} {Imperial College London} (\bibinfo {year}
  {2014})\BibitemShut {NoStop}%
\bibitem [{\citenamefont {Born}\ and\ \citenamefont {Wolf}(1998)}]{Born1998}%
  \BibitemOpen
  \bibfield  {author} {\bibinfo {author} {\bibfnamefont {M.}~\bibnamefont
  {Born}}\ and\ \bibinfo {author} {\bibfnamefont {E.}~\bibnamefont {Wolf}},\
  }\href@noop {} {\emph {\bibinfo {title} {Principles of {Optics}}}}\ (\bibinfo
   {publisher} {{Pergamon Press, Oxford}},\ \bibinfo {year} {1998})\BibitemShut
  {NoStop}%
\end{thebibliography}%

	\onecolumngrid
	\clearpage
	
	\renewcommand{\theequation}{S\arabic{equation}}
	\renewcommand{\thefigure}{S\arabic{figure}}
	\setcounter{equation}{0}
	\setcounter{figure}{0}
	\setcounter{page}{1}

	\section{Supplemental Material for: ``Spatial and Temporal Coherence in Strongly Coupled Plasmonic Bose--Einstein Condensates''}
	\twocolumngrid
	
	\section{Samples}
	
\begin{figure}[b]
		\vspace{-12pt}
		\centering
		\includegraphics[width=0.9\columnwidth]{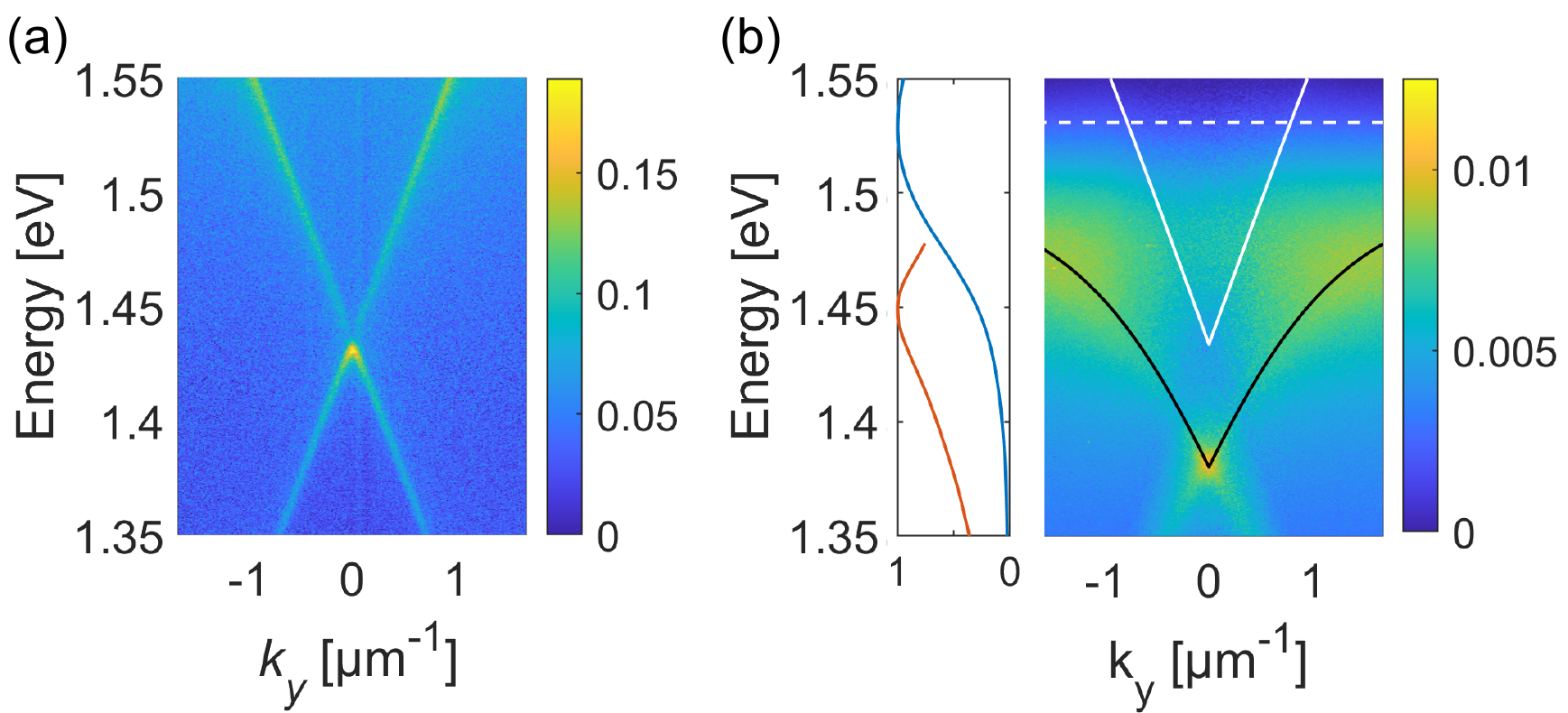}
		\vspace{-5pt}
		\caption{Dispersion relations of the sample with and without dye. (a) Extinction spectrum of the nanoparticle array obtained by transmission measurement. (b) Reflection measurement of the nanoparticle array with 80~mM solution of IR-792 dye molecules. The dispersion of the bare array from (a) is shown by the white solid line. The black solid line is a fit of the reflection maxima to the coupled modes model. The white dashed line points out the absorption maximum of the dye, whose normalized absorption (blue) and emission (red) spectra are shown on the left~\cite{Vakevainen2020}.}
		\label{figureSXX_dispersions}
	\end{figure}

	The nanoparticle arrays are fabricated on a glass substrate with electron beam lithography (EBL). A 2~nm adhesion layer of titanium and a 50~nm layer of gold are evaporated on a patterned poly(methyl methacrylate) layer on glass, followed by lift-off in acetone. Gold was chosen as the nanoparticle material because it is resistant to oxidation and has low losses in the near-infrared wavelengths where the experiments are conducted~\cite{johnson_optical_1972}. The arrays have asymmetric periodicities ($p_y=568$~nm, $p_x=618$~nm) and the cylindrical gold nanoparticles have a diameter of 105~nm and height of 50~nm. The asymmetric periods separate the dispersion of orthogonal polarizations, simplifying the data interpretation~\cite{Vakevainen2020}. The dispersion relation of the transverse electric (TE) SLR mode obtained by transmission measurement is shown in Fig.~\ref{figureSXX_dispersions}(a). The nanoparticle arrays are covered with fluorescent dye IR-792 (80~mM concentration) in a solution of 1:2 (dimethyl sulfoxide) : (benzyl alcohol). A large molecule reservoir (0.8~mm thick layer) helps the samples to sustain measurements for a long period of time because the dye can replenish between consecutive measurements. After adding the dye molecules, we observe the typical signatures of strong coupling in the dispersion (reflection measurement, see Fig.~\ref{figureSXX_dispersions}(b)): avoided crossing between the SLR mode and the dye absorption, and consequently a red shift of the dispersion band edge. The band edge in the uncoupled array is at 1.431~eV while in the coupled samples at 1.382~eV. A fit of the reflection maxima to the coupled modes model~\cite{torma_strong_2015} gives a Rabi splitting of 180~meV.

	The dye molecules are excited by a pulsed Ti:sapphire laser (50~fs, 1~kHz, 800~nm/1.55~eV), the spot is cropped by an iris such that it has a diameter larger than the arrays and nearly flat-top intensity profile. The pump is horizontally polarized ($x$ direction of the nanoparticle array). Even though the pump does not directly couple to the lattice modes (a pump beam at 800~nm and normal incidence is off-resonant with the SLR modes of our sample), a small spectral overlap of the pump and the single particle resonance (which is broad) excites the nanoparticles and causes the nanoparticle charge oscillations to be mainly polarized in $x$. Small nanoparticles act as dipole antennae, radiating mostly in the direction perpendicular to their polarization axis. This corresponds to an SLR mode with propagation along the $y$-axis of the lattice. The small pump-caused $x$-polarized excitation in the nanoparticles stimulates $x$-polarized emission from the molecules, and thereby the thermalization is triggered to an SLR mode propagating along the $y$-axis of the lattice (Fig.~\ref{figureSXX_dispersions}).

	\begin{figure}[b]
		\vspace{-12pt}
		\centering
		\includegraphics[width=\columnwidth]{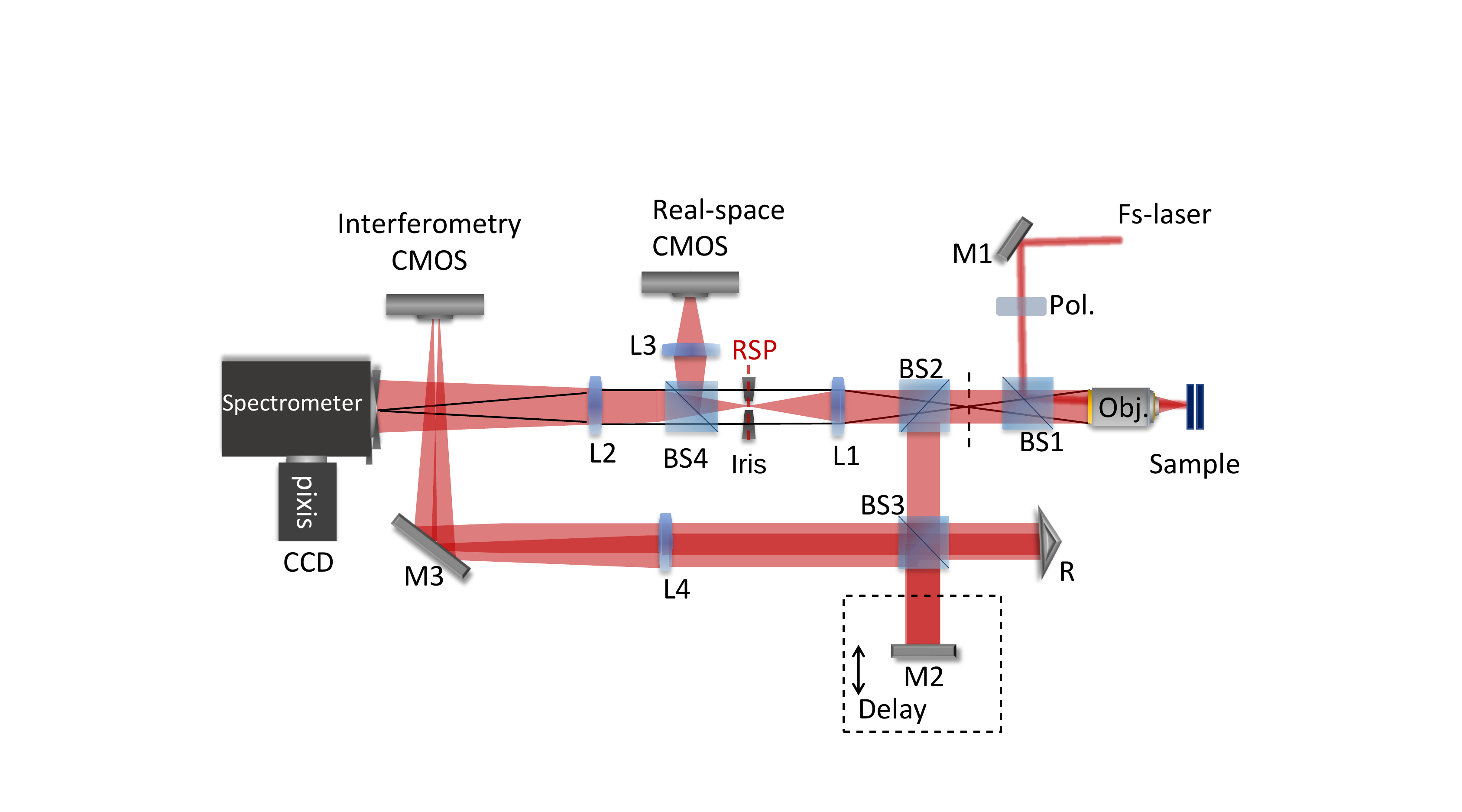}
		\vspace{-5pt}
		\caption{Measurement setup. The sample was excited at a normal incidence by a 50~fs, 800~nm pulse (1.55~eV with full width at half maximum of 0.04~eV). The mirror (M) 1 and beam splitter (BS) 1 were used to direct the excitation through the objective lens (Nikon 10x NA~0.3). A polarizer was used to maintain linear polarization of the excitation. The same objective was used to collect the sample luminescence, and BS2 allowed us to simultaneously perform spatial and temporal correlation measurements with a Michelson interferometer in a retroreflector configuration (Fig.~1). The Michelson interferometer accommodates a delay mirror (M2) and the retroreflector (R). The inverted and non-inverted images are focused to a CMOS camera with the $f=450$~mm bi-convex lens (L4). To spatially crop the sample luminescence, an iris was placed at the real space image plane (RSP) after the $f=200$~mm tube lens, L1. The beam splitter BS4 was used to allow the collection of the real space image with L3 and the angle-resolved luminescence spectrum with L2. }
		\label{fig:measurement_setup}
	\end{figure}

\begin{figure}[b]
		\vspace{-12pt}
		\centering
		\includegraphics[width=0.85\columnwidth]{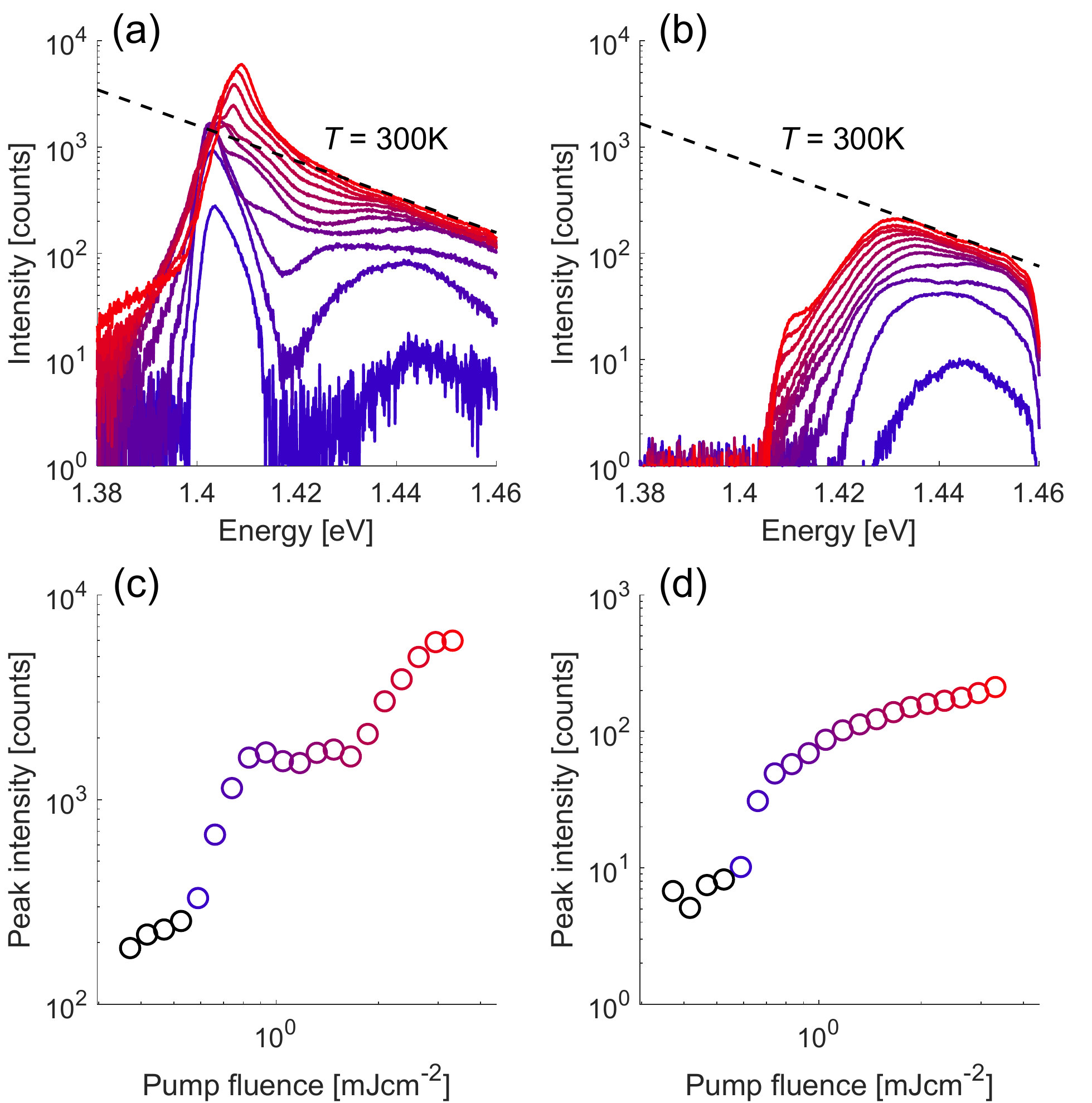}
		\vspace{-5pt}
		\caption{Luminescence spectra as a function of pump fluence. Luminescence spectrum for (a) the whole spectrum and (b) the thermal cloud as a function of pump fluence. The pump fluences are color-coded from low (blue) to high (red) respective to (c) and (d), which show the threshold curves for the whole spectrum and the thermal cloud, respectively. Notably, the population of the thermal tail increases linearly as a function of pump fluence after the first threshold. At the second threshold this increase saturates and the population starts to accumulate to a peak around the band edge, in accordance with the BEC mechanism.}
		\label{figureSXX_spectra}
	\end{figure}

	As observed in our earlier work~\cite{Vakevainen2020}, pumping degrades strong coupling up to some degree, manifested as blue shift of the dispersion.
	At the pump fluence where BEC is formed (3.31~mJcm$^{-2}$) the band edge emission is blue shifted to 1.409~eV (Fig.~\ref{figureSXX_spectra}). According to the coupled modes model, this corresponds to a Rabi splitting of 115~meV. This still fulfills the strong coupling condition $\Omega_\mathrm{R} > (\gamma_{\mathrm{dye}} + \gamma_{\mathrm{SLR}})/2 \approx 50$~meV, confirming that the system remains at the strong coupling regime throughout the studied pump regime.
	
	\begin{figure*}
		\centering
		\includegraphics[width=0.7\textwidth]{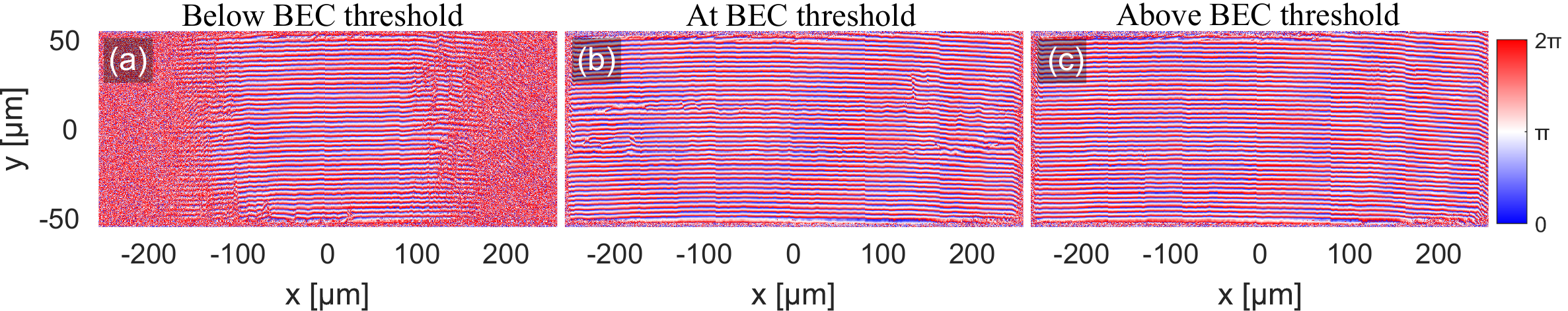}
		\vspace{-5pt}
		\caption{Maps of the phase of the interference fringes. Phase $\phi$ obtained by the analysis of interference fringes at three pump fluences (a) below (0.83~mJcm$^{-2}$), (b) at (1.66~mJcm$^{-2}$), and (c) above (3.31~mJcm$^{-2}$) the BEC threshold, respective to Fig.~2 of the main text.
		}.
		\label{figureSXX_phase}
		\vspace{-15pt}
	\end{figure*}
	
	\section{Transmission and reflection measurement}
	Dispersion of optical modes in the nanoparticle array are obtained by transmission measurement. The sample is illuminated with a white-light source (halogen lamp) from the back side of the sample, and the transmitted light is resolved by a spectrometer. The modes are visible as dips (peaks) in the angle-resolved transmission (extinction) spectrum, see Fig.~\ref{figureSXX_dispersions}(a). Modes of the sample with dye molecules are measured by reflection, because the thick layer of dye prevents transmission of light through the sample. The sample is illuminated from the front side with a halogen lamp, and the reflected light is resolved by the spectrometer. The modes are visible as peaks in the angle-resolved reflection spectrum, see Fig.~\ref{figureSXX_dispersions}(b).

	\section{Spatial coherence measurement}
	
	Spatial and temporal coherence is measured with a Michelson interferometer in a mirror-retroreflector configuration. The measurement setup, depicted in Fig.~\ref{fig:measurement_setup}, allows measuring simultaneously with the Michelson interferometer and a spectrometer in order to record the angle-resolved spectrum of the sample luminescence. Monitoring the luminescence spectrum is important to verify the stability of the condensate throughout the temporal and spatial coherence measurements. 
	
	In the Michelson interferometer, the sample luminescence is split to two arms of the interferometer by a 50/50 cube beam splitter. One of the arms is equipped with a retroreflector, while the other arm is with a flat mirror on a motorized linear stage. The location of the stage therefore sets the time delay between the two arms. The retroreflector inverts the real space image centrosymmetrically i.e. in both $x$ and $y$ directions of the array. The inverted and non-inverted images are overlapped at a complementary metal–oxide–semiconductor (CMOS) camera and the combined image shows interference fringes when the emission from the sample is coherent. 
	
	We measure a series of interferograms at fixed intervals around the zero time delay, $\tau=0$, over approximately three cycles of light frequency oscillation. Three oscillations account for $\sim9$~fs (at 880~nm), which is divided to 21 delay steps providing a stack of interferograms. The raw interferograms, with intensity denoted as $I_\mathrm{raw}$, are normalized with images taken separately from each arm of the interferometer ($I_1$ and $I_2$). The interferogram with normalized intensity is $I=(I_\mathrm{raw}-I_1-I_2)/(2\sqrt{I_1I_2})$~\cite{daskalakis_room-temperature_2014}. 
	The fringe contrast $C$ is directly proportional to the absolute value of the first-order correlation function $|g^{(1)} \left(\textbf{r}, -\textbf{r}; \tau \right)|$~\cite{Born1998}:
	\begin{align}
		\label{eq:g1}
		C\left(\textbf{r},\tau \right) = \frac{I_\mathrm{max}-I_\mathrm{min}}{I_\mathrm{max}+I_\mathrm{min}} = \frac{2\sqrt{I\left( \textbf{r}  \right)I\left( -\textbf{r}  \right)}}{I\left( \textbf{r} \right)+I\left( -\textbf{r}  \right)}|g^{(1)} \left(\textbf{r}, -\textbf{r};\tau \right)|.
	\end{align}
	Background counts, obtained by blocking the view of the camera, are subtracted from all the images before analysis. All images taken by the CMOS camera are integrated over 490~ms i.e. 490 excitation pulses at the 1~kHz repetition rate.
	
	\begin{figure}[b]
		\centering
		\includegraphics[width=0.95\columnwidth]{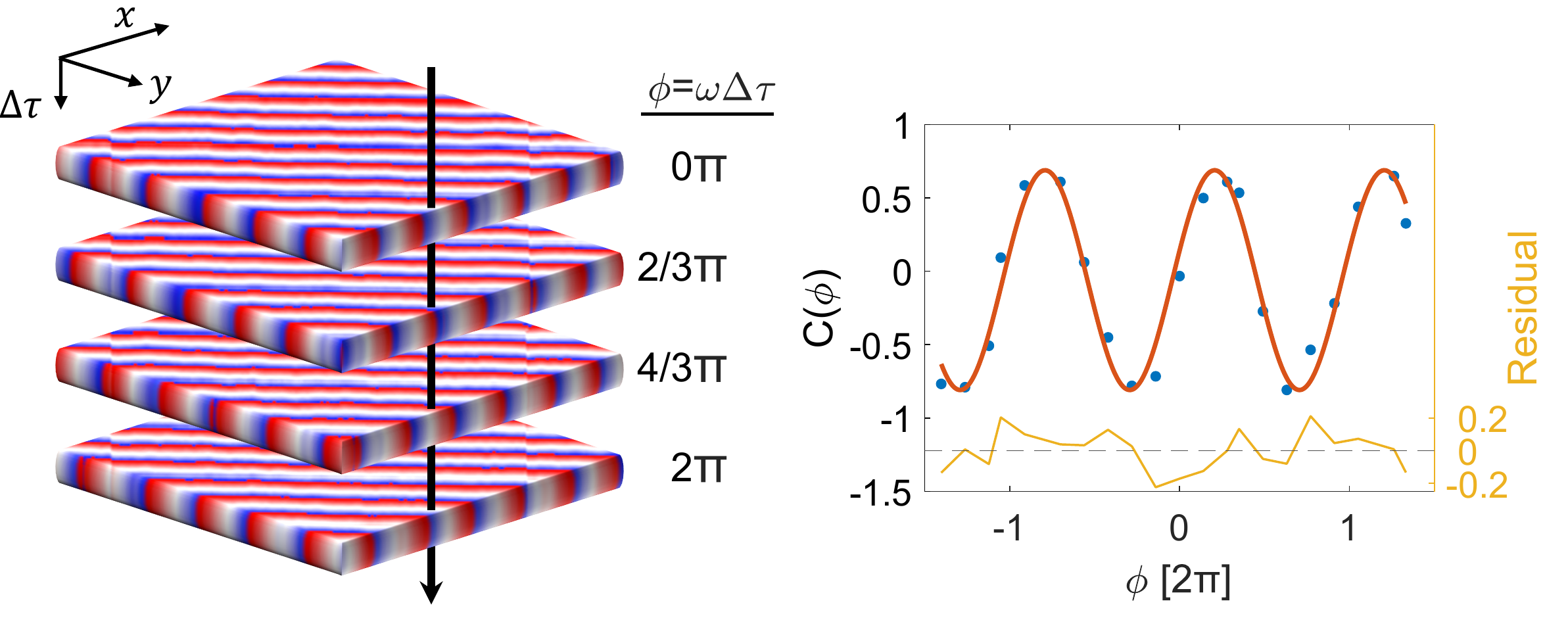}
		\vspace{-10pt}
		\caption{Schematic illustration of the analysis of interference fringes. A sinusoidal function is fit through each pixel of the stack of 21 normalized interferograms (left panel). An example fit (red line) with 21 data points (blue dots) is shown in the right panel. Three periods of light frequency oscillation account for approximately $\sim9$~fs.}
		\label{figureSXX_sinfit}
	\end{figure}
	
	\begin{figure}
		\centering
		\includegraphics[width=0.7\columnwidth]{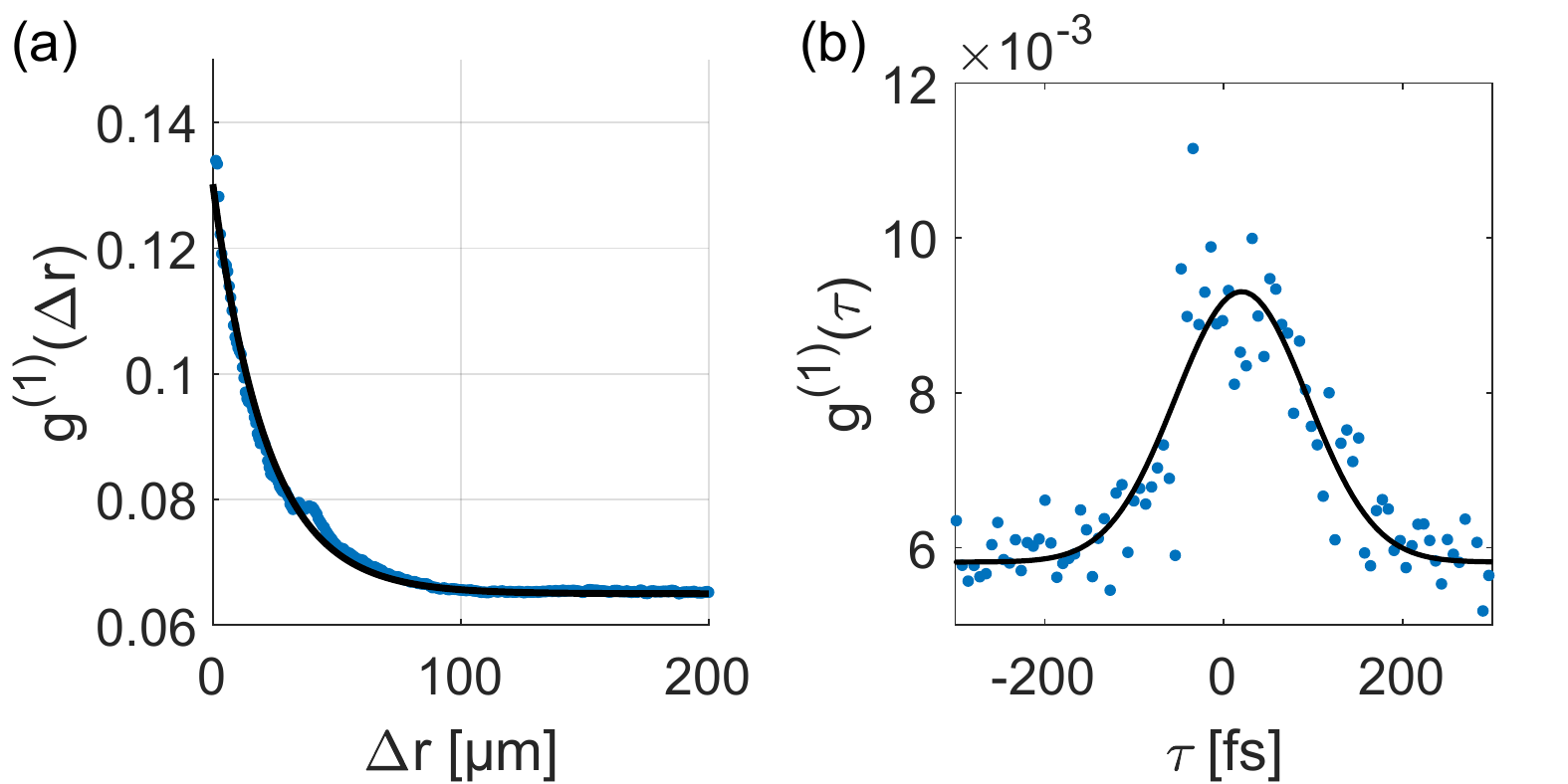}
		\vspace{-5pt}
		\caption{Spatial and temporal coherence of the polaritons below the first threshold. (a) Spatial coherence and fit to an exponential function that gives a decay length of \SI{22}{\micro m}. (b) Temporal coherence and fit to a Gaussian function with half-width at 1/e decay of 104~fs. The results were obtained at the lowest pump fluence which gave a detectable signal ($\sim 0.3$~mJcm$^{-2}$), below the first (lasing) threshold.}
		\label{figureSXX_belowth}
	\end{figure}

	\begin{figure*}
		\centering
		\includegraphics[width=0.7\textwidth]{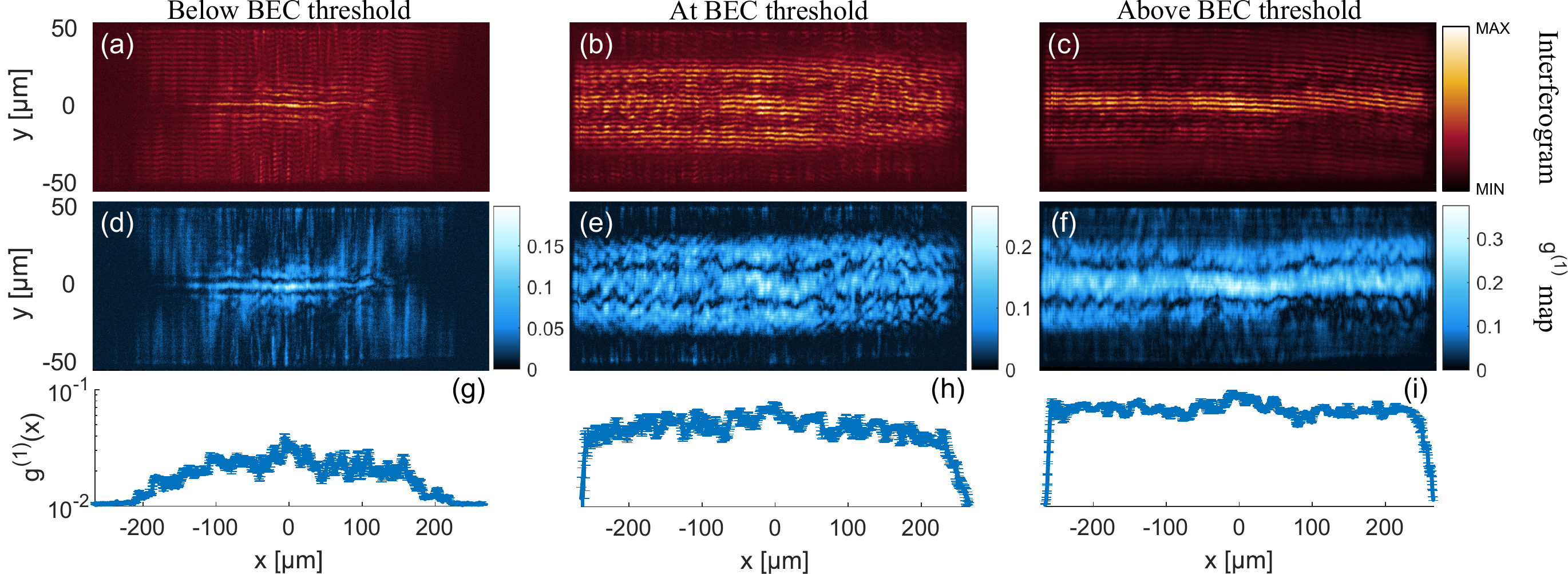}
		\vspace{-5pt}
		\caption{Interferograms and spatial correlation maps for the thermal cloud. Interferograms at three pump fluences (a) below (0.83~mJcm$^{-2}$), (b) at (1.66~mJcm$^{-2}$), and (c) above (3.31~mJcm$^{-2}$) the BEC threshold. (d-f) Maps of $g^{(1)}$ for the corresponding pump fluences. (g-i) Average $g^{(1)}(x)$ taken over the $y$-axis in (d-f). The error bars represent the standard deviation of three measurements. Due to low number of counts, the thermal cloud data was analyzed without subtracting the background counts. Subtracting the background would lead to large negative values in the normalization procedure. 
			Consequently, the absolute values of $g^{(1)}$ here are not correctly normalized between $[0,1]$. To separate the thermal cloud from the sample luminescence, we used two band pass filters centered at 850~nm (1.458~eV) with a bandwidth of 40~nm. We note that the filters let some emission through at the band-edge energy (see Fig.~\ref{figureSXX_spectra} for the photoluminescence spectra), which may contribute to the coherence.
		}
		\label{Fig_thermaltail}
		\label{figureS2}
		\vspace{-5pt}
	\end{figure*}

	\begin{figure}[b]
		\centering
		\includegraphics[width=0.95\columnwidth]{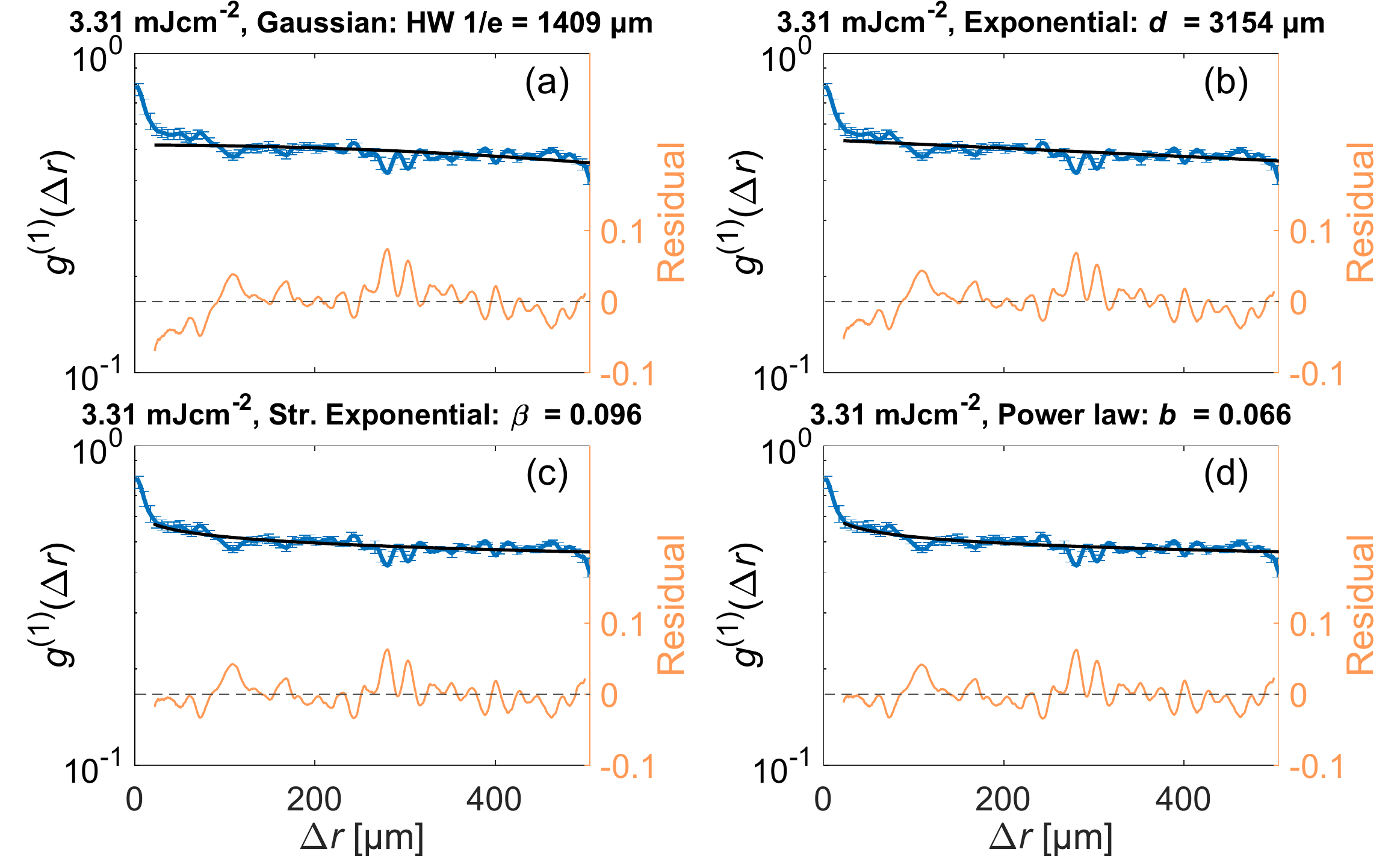}
		\vspace{-10pt}
		\caption{Fits of the spatial correlations to all four functions. Non-linear least squares fit of measured spatial correlations at pump fluence 3.31~mJcm$^{-2}$ to (a) Gaussian, (b) exponential, (c) stretched exponential, and (d) power-law function. }
		\label{figureSXX_allfits_bec}
	\end{figure}
	
	To obtain the fringe contrast, we fit a sinusoidal function, $ C \sin (\omega \tau + \phi) + C_0$, through each pixel of the stack of normalized interferograms, providing the amplitude $C$ and phase $\phi$ of the fringes (see Fig.~\ref{figureSXX_sinfit} for illustration). The maps of the phase $\phi$ of the interference fringes are shown in Fig.~\ref{figureSXX_phase}. The phase reflects the ordering of correlations and can be used to confirm that the contrast has been extracted reliably by the fits. Under optimal conditions and perfectly symmetric condensate with $I(\textbf{r})=I(-\textbf{r})$, the fringe contrast $C$ is equal to $|g^{(1)} \left(\textbf{r}, -\textbf{r}; \tau \right)|$. This is rarely the case in experiments, hence the maximum $|g^{(1)}|<1$. In addition to spatial intensity variations, the obtained $g^{(1)}$ values are reduced by time-integrating (mechanical vibrations of the setup may slightly change the path lengths of the interferometer arms etc.) and by imperfections in the optics, such as aberration~\cite{Born1998}.
	

	

	When performing the experiments, the delay stage position that yields the highest fringe contrast indicates the $\tau=0$ position. The $g^{(1)}\left(\textbf{r}, -\textbf{r}; \tau=0 \right)$ is then obtained by collecting several scans of 21 interferograms around the estimated $\sim \tau=0$, and averaging the extracted $g^{(1)}$ values of a few scans closest to the $\tau=0$. The $g^{(1)}$ in Fig.~2(g-i) and Fig.~3(a-c) is an average of three scans within a range -10~fs to 10~fs at $\sim \tau=0$. 
	
	To analyze spatial correlations from the maps of $g^{(1)}$, we take average $g^{(1)}(x,y)$ on a circumference with increasing diameter $(\Delta r)$ starting from the autocorrelation point at the center of the array. The obtained $g^{(1)}(\Delta r)$ are fit to Gaussian, exponential, stretched exponential, and power-law functions, and the fit quality is assessed for each function. In order to study long-range correlations, a range between zero and the spatial coherence length of the polaritons in the uncondensed system, below the first threshold (\SI{22}{\micro m}, see Fig.~\ref{figureSXX_belowth}) is excluded from the fits. In addition to graphical evaluation, the fits are compared by their residuals and root-mean-square error (RMSE). 
	The RMSE of the fits in Fig.3(g) shows a trend that at low pump fluences the Gaussian model has the lowest RMSE, whereas at high pump fluences the power law and stretched exponential have the lowest RMSE. Closer inspection of the fits and residuals further suggests that the Gaussian and exponential functions do not fit as well above the BEC threshold (Fig.~\ref{figureSXX_allfits_bec}) as the stretched exponential and power law models.
	
	That the spatial coherence extends to such large distances over the lattice in $x$ direction is non-trivial considering the intrinsic anisotropy of the process. The pump excites mainly the molecules but also weakly (off-resonantly) the nanoparticles: the nanoparticle excitations follow the pump polarization and stimulate emission from the molecules to SLR modes of the same polarization (see Fig.~\ref{figureSXX_dispersions}). In the experiments reported in the manuscript, the pump polarized linearly in $x$ triggers propagation of high energy (high momentum $\mathbf{k}$) polaritons in the $y$ direction of the array. In the beginning of the thermalization process, $y$ direction is thus favoured. Nevertheless, the BEC that forms around the band edge ($\mathbf{k}=0$) shows coherence in both $x$ and $y$, in contrast to lasing at the first threshold which is coherent only in $y$. This qualitative change of spatial coherence from lasing to BEC was found already in our previous work~\cite{Vakevainen2020}, but for a sample of only \SI{100}{\micro m} size in $x$ (coherence decay laws were not studied there, limited by the small sample size). The present observations of coherence in the $x$ direction over a much longer distance corroborate the 2D nature of the BEC.

	We note that control experiments with a random array sample did not show any coherence (see Fig.~\ref{figureSXX_random}). The array had a random particle distribution and the same particle dimensions and density as the arrays in the main experiments. This shows that pumping of the molecules, in the presence of randomly distributed nanoparticles, does not create coherence. The SLR modes, arising from periodicity, are needed for the BEC and coherence formation.
	
	The data acquisition and analysis are demanding; the sample must remain stable throughout numerous measurements at pump fluences above the BEC threshold, and each measurement produces 21 interferograms which all contain approximately $400\times2000$ pixels where the fits are performed. To reduce the computational cost, we deployed parallel computing for the non-linear least-squares fits. In turn, compared to other standard methods, such as Fourier analysis of spatial frequencies, this method provides much more robust results.
	
	\section{Temporal coherence measurement}
	
\begin{figure}[h!]
		\centering
		\includegraphics[width=1\columnwidth]{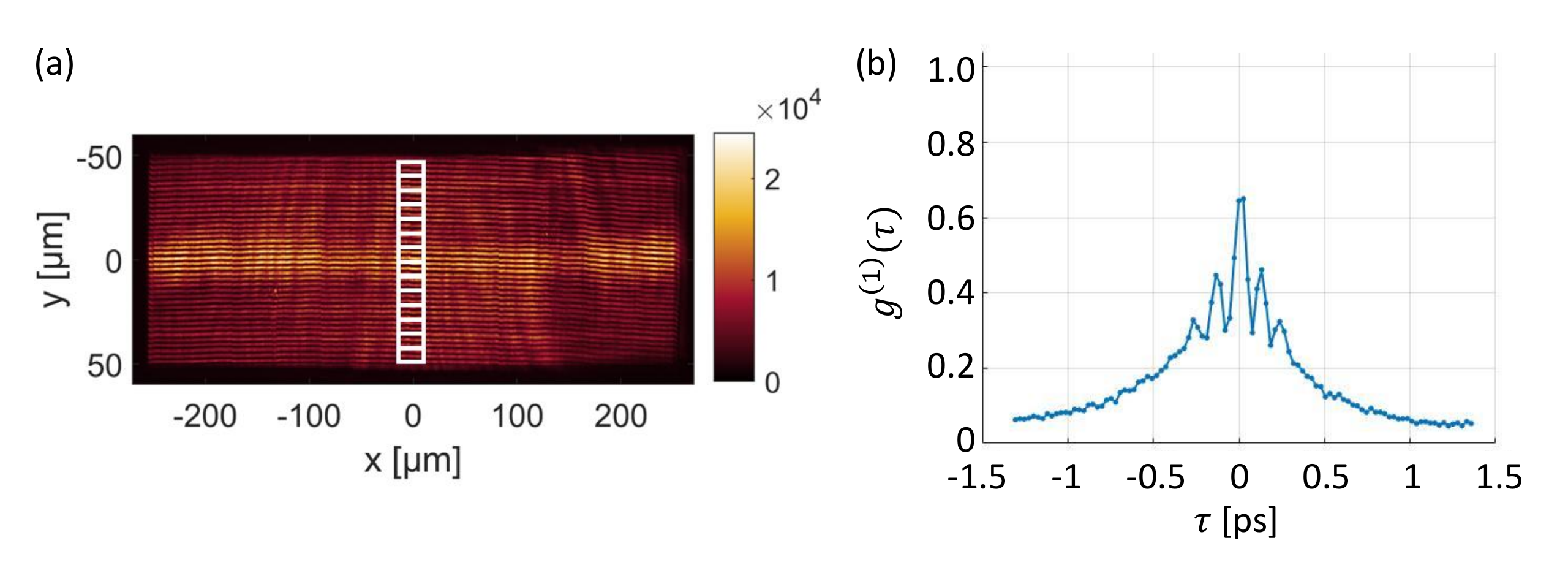}
		\vspace{-20pt}
		\caption{Spatial locations for temporal correlation analysis. (a) The white rectangles ($22\times$\SI{11}{\micro m}) illustrate the spatial locations from which the temporal correlations were extracted around $x=0$. (b) An example fringe contrast data from one rectangle, displaying interference side peaks that arise from counter-propagating polaritons. }
		\label{figureSXX_temporalint}
		\vspace{-10pt}
	\end{figure}

\begin{figure}[h!]
		\centering
		\includegraphics[width=0.6\columnwidth]{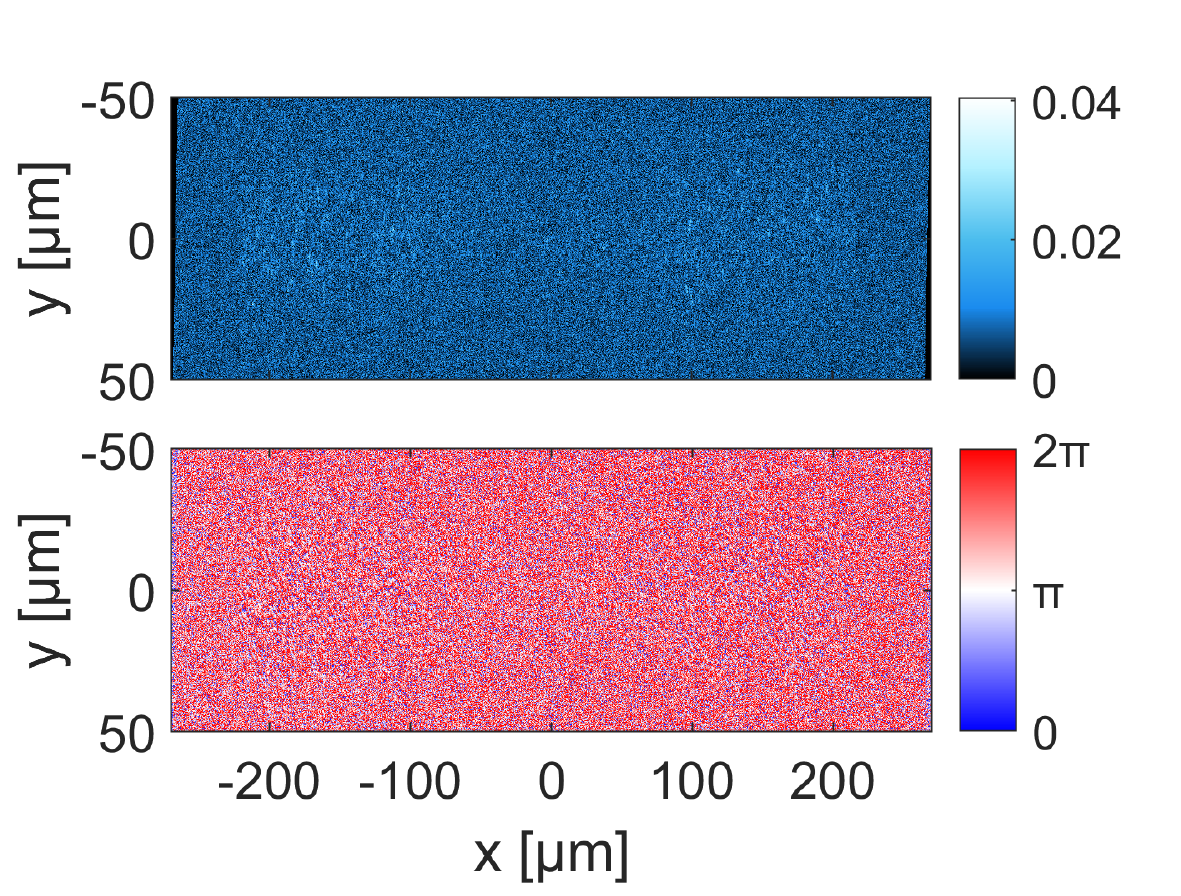}
		\vspace{-5pt}
		\caption{Spatial coherence of a random array sample. Spatial coherence measured from a sample where the nanoparticles are randomly distributed, with the same particle shape, size, and density as in the BEC samples. The array is overlaid with 80~mM dye solution. The results are shown for the pump fluence that corresponds to the BEC regime in the main text ($3.31$~mJcm$^{-2}$). } 
		\label{figureSXX_random}
	\end{figure}

	For temporal coherence, interferograms were recorded over delays ranging from -1.25~ps to 1.25~ps at fixed (30~fs) intervals. 
	Here, contrary to the spatial coherence analysis, we applied a simpler method to extract the fringe contrast by fitting a sinusoidal function directly to the normalized interferograms at each delay. The fringes were analyzed from \SI{22}{\micro m} wide rectangles around $x=0$, and the obtained $g^{(1)}(\tau)$ was averaged over all locations in $y$ as depicted in Fig.~\ref{figureSXX_temporalint}(a), see also the explanation below. The interference fringes inside the rectangular regions were fit to a sinusoidal function, which gives the amplitude 
	of the fringes. The contrast is directly proportional to the first-order correlation function $g^{(1)}(\tau)$, analogous to the spatial coherence analysis. In all data shown here and in the manuscript there was no scaling factor between the $g^{(1)}(\tau)$ and the contrast; below, in Fig.~\ref{pulselength} we discuss the effect of scaling that comes from finite pulse length. For the fits, we take the average of $g^{(1)}(\tau)$ over negative and positive delays $\pm \tau$ in order to get $g^{(1)}(|\tau|)$. The measured $g^{(1)}(|\tau|)$ is fit to Gaussian, exponential, stretched exponential, and power-law functions. We use again the same fit range for all functions and pump fluences. We exclude from the fits a range between zero and the temporal coherence of the polaritons in the uncondensed system, below the first threshold ($\tau=104$~fs, see Fig.~\ref{figureSXX_belowth}).

	When extracting the temporal coherence from very small spatial regions, we observed side peaks (see Fig.~\ref{figureSXX_temporalint}(b) for an example) which arise from interference between counter-propagating polaritons (i.e.~not reflections) along $\pm y$ of the lattice~\cite{Vakevainen2020}. As explained above and in Ref.~\cite{Vakevainen2020}, the pump polarization triggers the thermalization to start along the modes where excitations propagate along the $y$ direction. For the data used in the analysis of temporal correlation decay, we averaged over all the locations along the $y$-axis so that the interference peaks do not compromise the non-linear least squares fitting. 
	In case of symmetric/isotropic polariton and photon condensates (such as in Refs.~\cite{caputo_topological_2018, marelic_spatiotemporal_2016}), temporal coherence is often obtained at the center of the sample.

	The temporal coherence may be underestimated due to the finite length of the output pulses. When the output pulse is split in two in the Michelson interferometer and one is delayed, the parts of the two pulses that overlap at the detector actually diminish exponentially with the delay (assuming Gaussian pulse shape): this is illustrated in the schematic in Fig.~\ref{pulselength}(a). The interference fringes can occur only in the overlap region, while the total intensity of both pulses is collected by the camera: the non-interfering part of the two pulses adds as a background and reduces the fringe contrast, and thus the $g^{(1)}$ value if no normalization is made to take into account this effect (such underestimation of coherence is discussed also in previous literature, e.g., Ref.~\cite{de_giorgi_interaction_2018}).

	A simple estimate for this effect can be obtained by normalizing the measured fringe visibility to a factor that takes into account the temporal intensity decay of the pulses. As $g^{(1)}(\tau)$ tells about the coherence between the field amplitudes $E(t)$ and $E(t+\tau)$, the intensity of the recorded interferogram as a function of time $t$ and delay $\tau$ can we expressed as
	\begin{align}
		&\int_{-\infty}^{\infty} |E(t) + E(t-\tau)|^2 \text{dt} \\ &=\int_{-\infty}^{\infty} I(t)+I(t-\tau) +2g^{(1)}(\tau)\sqrt{I(t) I(t-\tau)}\text{cos}(\phi) \text{dt} \nonumber,
	\end{align}
	where $\phi$ is the phase difference between the field amplitudes and $I(t) = |E(t)|^2$ is the intensity. For time-independent intensity, $\int_{-\infty}^{\infty} \left[ I(t)+I(t-\tau)\right] \text{dt} = \int_{-\infty}^{\infty} 2\sqrt{I(t) I(t-\tau)} \text{dt} $ and the fringe contrast directly gives $g^{(1)}(\tau)$, however, for a pulsed source the contrast may underestimate $g^{(1)}(\tau)$ as discussed above. Thus the measured visibility could be normalized to the scaling factor
	\begin{align}
		F(\tau)=\frac{2\int_{-\infty}^{\infty}\sqrt{I(t) I(t-\tau)}\text{dt}} {\int_{-\infty}^{\infty}{\left[ I(t)+I(t-\tau)\right]}\text{dt}},
	\end{align}
	which for a Gaussian pulse shape, $I(t) = \text{exp} (-t^2/\sigma^2)/(\sigma\sqrt{\pi})$, becomes $F(\tau)=\text{exp}(-\tau^2/(4\sigma^2))$.
	
	We do not know the shape or the duration of the output pulse, although based on earlier pump-probe measurements in similar (but not identical) systems~\cite{hakala_bose-einstein_2018,daskalakis_ultrafast_2018}, it could be of the order of a few picoseconds. Therefore, we use a Gaussian pulse shape with several pulse lengths from 0.5 to 4~ps for calculating the scaling factor. The result of the scaling is shown in Fig.~\ref{pulselength}b. For pulses of 2--4 ps, the $g^{(1)}$ indeed is increased overall, while still showing a decay as a function of time. For shorter pulses, however, the scaled coherence has an up-turn which is clearly unphysical. Too short pulses simply cannot provide the observed coherence decay, as their overlap would decrease much faster than that. This indicates that the duration of the output pulses is about 2 ps or longer.

	\begin{figure}[h!]
		\centering
		\includegraphics[width=0.6\columnwidth]{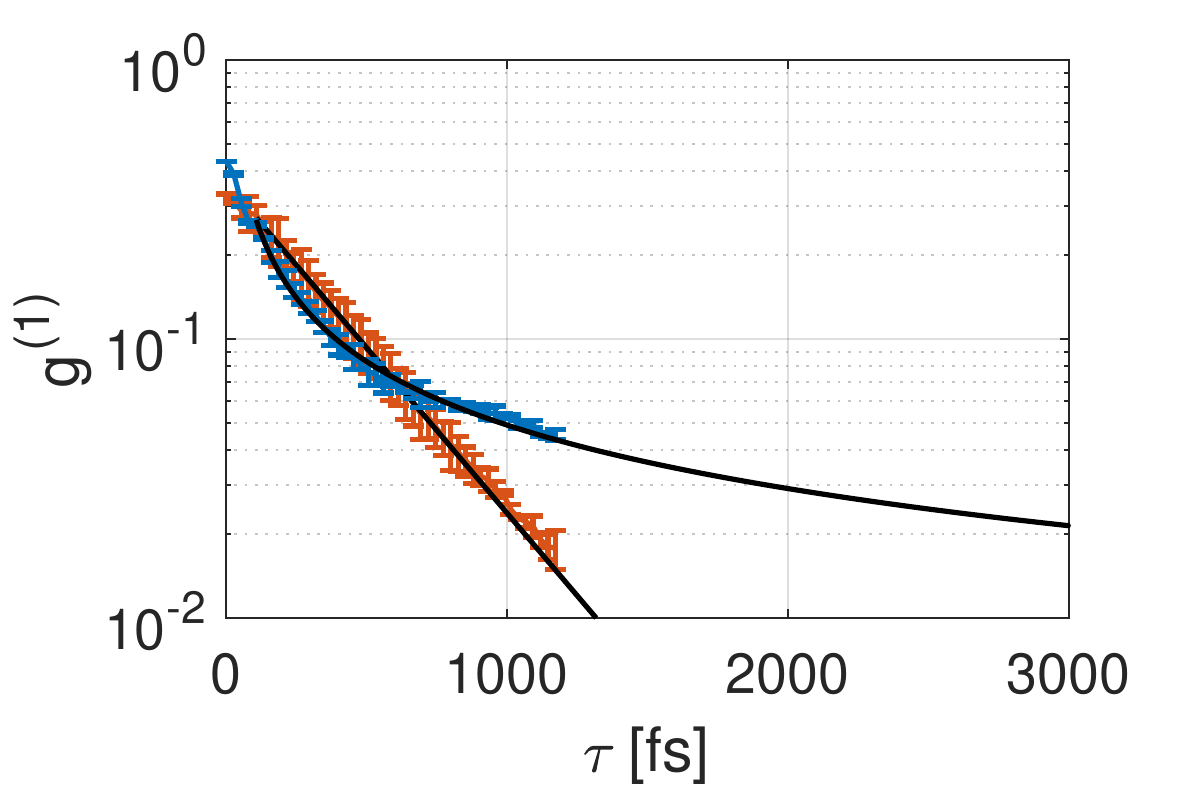}
		\caption{Comparison of the temporal coherence decay models in the polariton lasing (red) and BEC (blue) cases. The respective pump fluences are 0.83~mJcm$^{-2}$ and 3.31~mJcm$^{-2}$. The lasing case is fit by an exponential and the BEC case by a power law function.}
	\end{figure}
	
	\begin{figure}[h!]
		\centering
		\includegraphics[width=0.90\columnwidth]{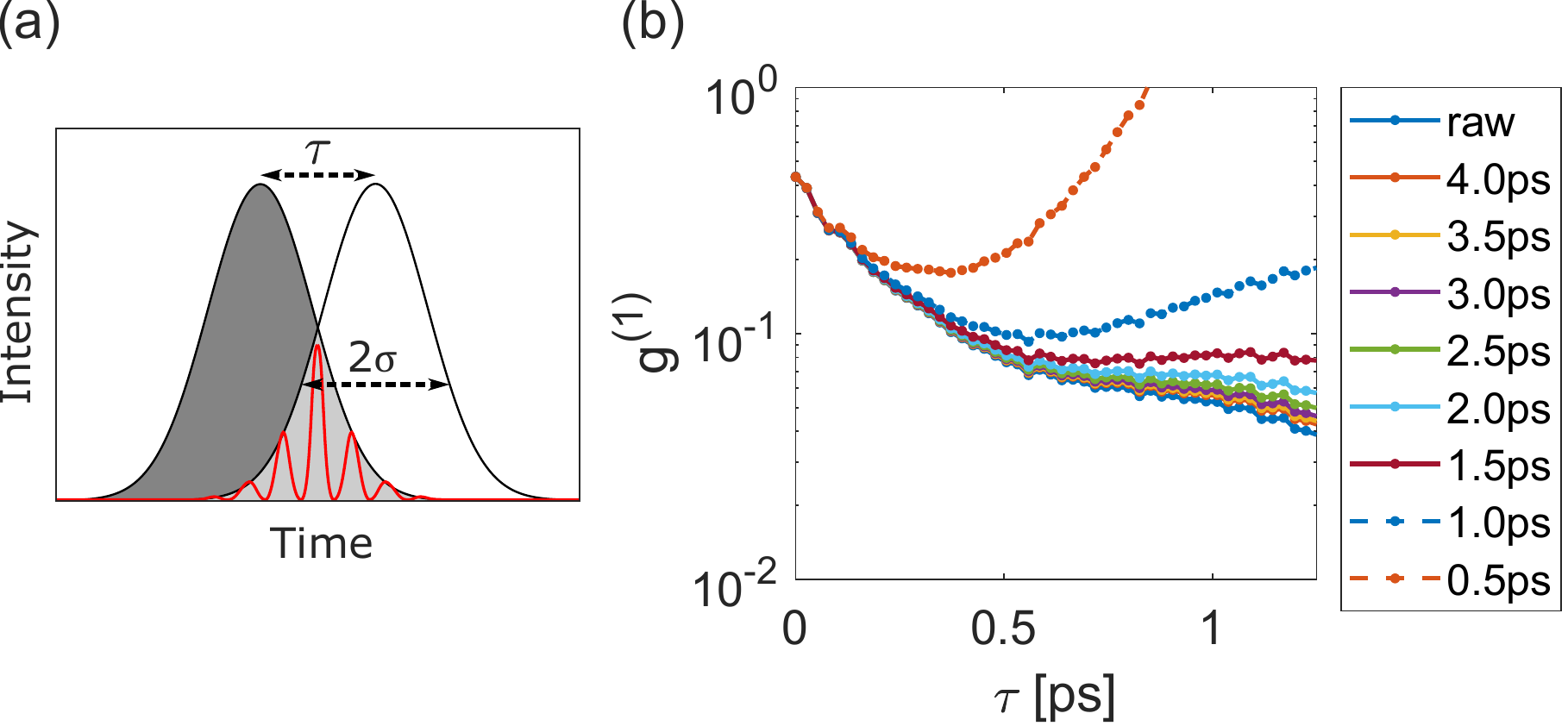}
		\caption{Effect of pulse duration on the temporal coherence. (a) Illustration of how only the overlap area of two time-separated pulses contributes to their interference. (b) The obtained $g^{(1)}(\tau)$ values for the BEC case, when divided by a scaling factor, for different output pulse durations. The output pulse is assumed to be a Gaussian, $I(t) = \text{exp} (-t^2/\sigma^2)/(\sigma\sqrt{\pi})$. The duration $2\sigma$ of the output pulse is indicated in the legend, as well as the unscaled (raw) data. The result shows that pulse durations below $2$~ps lead to increasing $g^{(1)}(\tau)$ as a function of $\tau$ which is unphysical and indicates that the pulse duration in our measurements is likely more than $2$~ps.}
		\label{pulselength}
	\end{figure}
	
	In the future, when comparing a theoretically obtained decay exponent to that obtained from our data, it is important to keep in mind this possible underestimation of temporal coherence due to finite pulse lengths. To determine precise values of decay exponents, one should have access to the actual pulse shape and duration.
	
\end{document}